# Before-after safety analysis of a shared space implementation

Federico Orsini[a,b]

Mariana Batista[c]

Bernhard Friedrich[c]

Massimiliano Gastaldi[a,b,d]

Riccardo Rossi[a,b]

[a]*Department of Civil, Environmental and Architectural Engineering, University of Padua, Italy*
[b]*MoBe – Mobility and Behavior Research Center, University of Padua, Italy*
[c]*Institute of Transportation and Urban Engineering, Technische Universität Braunschweig, Germany*
[d]*Department of General Psychology, University of Padua, Italy*

**To cite this article**:

Orsini, F., Batista, M., Friedrich, B., Gastaldi, M., & Rossi, R. (2023). Before-after safety analysis of a shared space implementation. *Case Studies on Transport Policy*, 101021.

**Link to the article:**

https://doi.org/10.1016/j.cstp.2023.101021

# Before-after safety analysis of a shared space implementation

## Abstract

Shared spaces aim to reduce the dominance of motor vehicles by promoting pedestrian/cyclist activity and minimizing segregation between road users. Despite the intended scope to improve the safety of vulnerable road users, only few works in the literature focused on before-after safety evaluations, mainly analyzing changes in users' trajectories and speeds, traffic volumes, and conflict counts, which, while useful, cannot univocally quantify road safety.

Here, we propose a more advanced methodology, based on surrogate measures of safety and Extreme Value Theory, to assess road safety before and after the implementation of a shared space. The aim is to produce a crash risk estimation in different scenarios, obtaining a quantitative and comprehensive indicator, useful to practitioners for evaluating the safety of urban design solutions.

A real-world case study illustrates the proposed procedure. Video data were collected on two separate days, before and after a shared space implementation, and were semi-automatically processed to extract road users' trajectories. Analysis of traffic volumes, trajectories, speeds and yield ratios allowed to understand the spatial behavior of road users in the two scenarios. Traffic conflicts, identified with an innovative surrogate measure of safety called "time to avoided collision point" (TTAC), were then used to estimate a Lomax distribution, and therefore to model the probabilistic relationship between conflicts and crashes, eventually retrieving a crash risk estimate. Results show that the analyzed shared space was able to significantly reduce the risk of crashes, and these findings are consistent with the observed changes in users speed and spatial behavior. The analyzed case study and its limitations were useful in highlighting the methodology main features and suggesting practical prescriptions for practitioners.

# 1. Introduction

The shared space concept proposes to integrate road users to regulate traffic and improve safety. The main assumption is that, by promoting barrier-free movement and minimizing traffic control, people become more alert to other road users, move slowly, and rely on interactions and negotiation-based movement to decide who has the right-of-way (Clarke, 2006; Hamilton-Baillie, 2008; Karndacharuk et al., 2014a; Pascucci et al., 2015; Rinke et al., 2017). Since shared spaces rely on interactions to function effectively, it challenges the well-established system of street segregation already designed to avoid conflicts and limit the contact between different road users (Hamilton-Baillie, 2008; Karndacharuk et al., 2014a).

As a result, shared spaces can be considered low-speed road environments that minimize segregation between road users. Beyond decreasing traffic speed and providing more space for pedestrians, the intention is to reduce the dominance of motor vehicles and reinforce the sense of place to further balance the functions of the street and promote it as a public space (Hamilton-Baillie, 2008; Karndacharuk et al., 2014a).

Despite the intended scope to improve the safety of vulnerable road users (VRU), only few works investigated the safety benefits of shared space implementations with before-after analysis. The lack of before-after studies on shared spaces was highlighted by Ruiz-Apilánez et al. (2017), and is also documented in the literature review of Section 2.1. In particular, the few existing before-after analyses of shared spaces focused on the study of changes road users' spatial behavior, which, while essential in understanding the effects and the implications of the infrastructural interventions, cannot be used to univocally determine changes in the safety of vulnerable road users. Some studies included analysis of traffic conflicts, which are more relevant for road safety; however, in the reviewed conflict-based approaches there are a few shortcomings: 1) in many cases conflicts are identified in a qualitative way, which adds an element of subjectivity and increases the time required for analysis, with negative consequences for practical applications, 2) they mostly just limit the analysis on a comparison between conflict rates in the before and after scenarios, without considering the more complex relationship that exists between conflict and crashes.

This study proposes a conflict-based methodology for before-after road safety evaluation of shared space implementations, filling a gap in the literature, which lacks procedures to univocally determine and quantify the safety benefits of these urban design solutions. In particular, the methodology presented here answers to the two main issues listed above by: 1) adopting a quantitative approach to detect and identify conflicts using a surrogate measure of safety called time-to-avoided-collision-point (TTAC); 2) applying an extreme value theory (EVT) approach to model the probabilistic relationship between conflicts and crashes.

The procedure is structured with practical operations in mind; therefore, it aims to be a theoretically-sound methodology, which should also be relatively easily applicable by practitioners.

A real-world case study is used to illustrate the application of the proposed methodology. It should be viewed as a test-bed to better understand the advantages and limitations of the methodology, rather than as an end itself. In this sense, the analyzed case study is a good way to showcase the main features of the procedure, but also, due to some of its limitations, to give relevant takeaways and suggest practical prescriptions.

The remainder of the paper is structured as follows: Section 2 presents a literature review of previous works dealing with before-after analysis of shared spaces, of surrogate measures of safety for vehicle-VRU conflicts, and of approaches for extracting crash probabilities from conflict data; Section 3 outlines the methodology used; Section 4 presents the case study and the data collection and preparation procedures; Section 5 illustrates the results of the before-after analysis; Section 6 discusses the main findings; Section 7 concludes the paper, summing up the main findings, and highlighting limitations and future perspectives.

# 2. Literature review

## 2.1 Before-after analysis of shared spaces

As anticipated in the Introduction, only few previous works investigated the safety of shared spaces with before-after analysis, mainly focusing on spatial behavior analysis and/or on qualitative evaluations of traffic conflicts.

Kaparias et al. (2015) compared road users' behavior before and after the redevelopment of Exhibition Road in London, aiming at evaluating pedestrians' confidence and vehicles' tolerance during interactions. Data were collected with video cameras, and the authors focused on interaction frequency, type, and severity. The before-after comparison was mainly descriptive but provided useful insights into factors affecting users' behavior in shared spaces, which can have relevant consequences in terms of safety.

A more directly safety-oriented before-after comparison of the same scenario was carried out by Kaparias et al. (2013), which adopted a conflict-based approach developed by Kaparias et al. (2010) and refined by Salamati et al. (2011), called pedestrian-vehicle conflict analysis (PVCA), which combines a quantitative measure of crash nearness, time-to-collision (TTC), with a qualitative evaluation of the interaction (e.g., severity of evasive action), categorizing conflicts into 4 grades of severity. The comparison of conflict rates, for each grade of severity, showed a general decrease after the redevelopment, indicating improvements in terms of safety. They later analyzed changes in gap acceptance behavior, observing higher comfort and confidence of pedestrians in their interaction with motor vehicles after the redevelopment (Kaparias et al., 2016).

Similar approaches were applied in a New Zealand case study, first investigating spatial-behavioral aspects (e.g., pedestrian trajectories, density, activity) (Karndacharuk et al., 2013), and then applying the PVCA for a before-after safety evaluation (Karndacharuk et al., 2014b).

More recently Lee and Kim (2019) analyzed nine "pedestrian priority street" projects in South Korea, which consisted of the application of special paving designs of various colors and patterns. The authors performed before and after video surveys and collected subjective users' evaluation with questionnaires. The main output of their work is that, in the analyzed scenarios, the design strategies were able to reduce vehicle speed and increase users' perception of safety, except in the areas where vehicular and pedestrian zones were still clearly distinguishable.

In the literature, there are some other examples of before-after analysis investigating the effect of other types of infrastructural interventions on the safety of vulnerable road users. Danaf et al. (2020) analyzed the installation of a new crosswalk in Beirut, Lebanon, assessing pedestrians' waiting time, accepted gap size, and approaching vehicle speed. Hirsch et al. (2022) evaluated the effect of several layout changes of an intersection in Auckland, New Zealand, which included interventions such as new road markings, and a new cycling lane; the authors focused their analysis on the interactions between users, defined according to a set of qualitative criteria. Navarro et al., (2022) assessed the conversion from minor-approach only stop to all-way stop intersections in the Montréal area, Canada, by comparing speed measures, yielding rates and users interactions (quantified by post-encroachment time, PET) before and after the intervention. Anciaes et al. (2020) studied yielding behavior at courtesy crossing, before and after the addition of a new element (i.e., stripes).

## 2.2 A surrogate measure of safety for vehicle-VRU conflicts

Most existing studies investigating the safety of VRUs adopted two popular surrogate measures of safety: time-to-collision (TTC) and post-encroachment time (PET).

TTC is a continuous variable. At time *t*, TTC is computed as the time until a collision assuming constant speed and trajectories of the users involved (Hayward J.C., 1972) (Figure 1a); the minimum value of TTC recorded during an interaction between road users is usually considered representative of the severity of the encounter, with low values of TTC identifying conflicts, and TTC = 0 representing a crash. TTC is considered one of the most popular surrogate measures of safety (Arun et al., 2021a, 2021b; Johnsson et al., 2018; Tarko, 2018a), and several modifications have been proposed (Nadimi et al., 2020). It has been applied in a wide range of studies investigating interactions between motor vehicles and pedestrians (Haque et al., 2021; Jiang et al., 2015; Kizawi and Borsos, 2021; Lenard et al., 2018; Zhang et al., 2020) or cyclists (Johnsson et al., 2021; Kovaceva et al., 2019; Orsini et al., 2022; Rossi et al., 2021; Zhao et al., 2021).

TTC can be computed in all those situations in which two users are on a collision course. In the case of car-following situations, this simply means that the follower should have a higher speed than the leader. On the other hand, when the trajectories of the users do not overlap, the condition on the speeds is no longer sufficient, as TTC can be computed only if the users involved are on course to occupy the trajectory intersection point at the same time. This is a serious limitation to the application of TTC, as many dangerous situations in which the users are just slightly off a collision course are completely missed by the measure; this has been observed for car-to-car conflicts (Gastaldi et al., 2021), and it is even more critical for conflicts involving pedestrians, who have smaller sizes than cars (Chen et al., 2017).

PET is calculated as the time difference between the instants in which the first user exits the conflict zone and the second one enters it (Allen et al., 1978) (Figure 1b); more generally it is defined as "the minimal delay of the first road user, which, if applied, will result in a collision course and a collision" (Laureshyn et al., 2010). Low PET values identify traffic conflicts, and PET = 0 represents a crash. It has been applied in many recent works on vehicle-VRU interactions (Chaudhari et al., 2021; Johnsson et al., 2021; Nasernejad et al., 2021; Vasudevan et al., 2022; Zhang et al., 2020; Zhu et al., 2021).

In contrast to TTC, PET is a directly observable measure that captures by how much time two road users missed each other. It describes what happened but gives no information on what could have happened; this is a serious limitation, as it prevents the possibility to apply a counterfactual analysis (Davis et al., 2011; Tarko, 2021, 2019). To better understand this concept, let us consider two road users on a collision course with a small TTC, and imagine that one of the users is able to avoid the collision, by braking to a full stop before continuing on its trajectory: in this case, the observed PET would be high and not representative of the risk involved in the encounter between the two vehicles. A possible solution is to consider what we could call a "counterfactual extension" of PET: that is, to assume, at time *t*, constant speed and trajectories of the road users and to compute the resulting expected PET. This surrogate measure of safety exists and is commonly referred to as time advantage (TAdv) (Laureshyn et al., 2010) (Figure 1c), although it is also known as "time difference to collision" (Zhang et al., 2012, 2011) and "relative time to collision" (Chen et al., 2017). Low values of TAdv are indicative of a potentially risky situation, and TAdv = 0 means that the two users are on a collision course. Recent examples of its application in VRU safety research exist (Gruden et al., 2022; Miani et al., 2022, 2021), but also this measure has a relevant shortcoming. As explained by Laureshyn et al. (2010), TAdv "is by itself not sufficient to describe the collision risk since it is also important to know how soon the encroachment will occur"; in practice, even if at time *t* TAdv is small, if the expected encroachment is far away, the users would have plenty of time to adjust their speed/trajectories and avoid each other without significant risk. The implication of this issue is that TAdv may identify traffic conflict situations that are not actually risky.

This problem can be solved with another surrogate measure of safety, derived from TAdv, which encloses information on the proximity to the encroachment. At time *t*, assuming constant speed and trajectory of the users, they will encroach with a certain TAdv; the difference between the instant at which the second vehicle is predicted to arrive at the avoided collision point and the current time *t* is a more safety-relevant measure, which was introduced by Laureshyn et al. (2010) and called $T_2$. Here, we will refer to it by using a more intuitive and citation-database-friendly name: time to the avoided collision point, TTAC (Figure 1c). TTAC reflects "the maximum time available to take evasive actions and alleviate the severity of the situation" (Laureshyn et al., 2017). The main advantage of TTAC is that it provides a "smooth" transfer between collision-course and crossing-course situations. As a matter of fact, it can be easily demonstrated that, when the road users are on a collision course (i.e., TAdv = 0), TTAC is equivalent to TTC; in other words, it includes all the qualities of TTC, without its main drawback, as it is possible to retrieve a value of TTAC even if the road users are bound to miss each other. As TTC, during a traffic encounter, TTAC is measured continuously and the minimum value of TTAC is usually considered as a single representative indicator of its severity. Despite this, only very few works applied TTAC for road safety evaluations (Borsos, 2021; Borsos et al., 2020; Miani et al., 2021). More details on its calculation are given in Section 3.2.

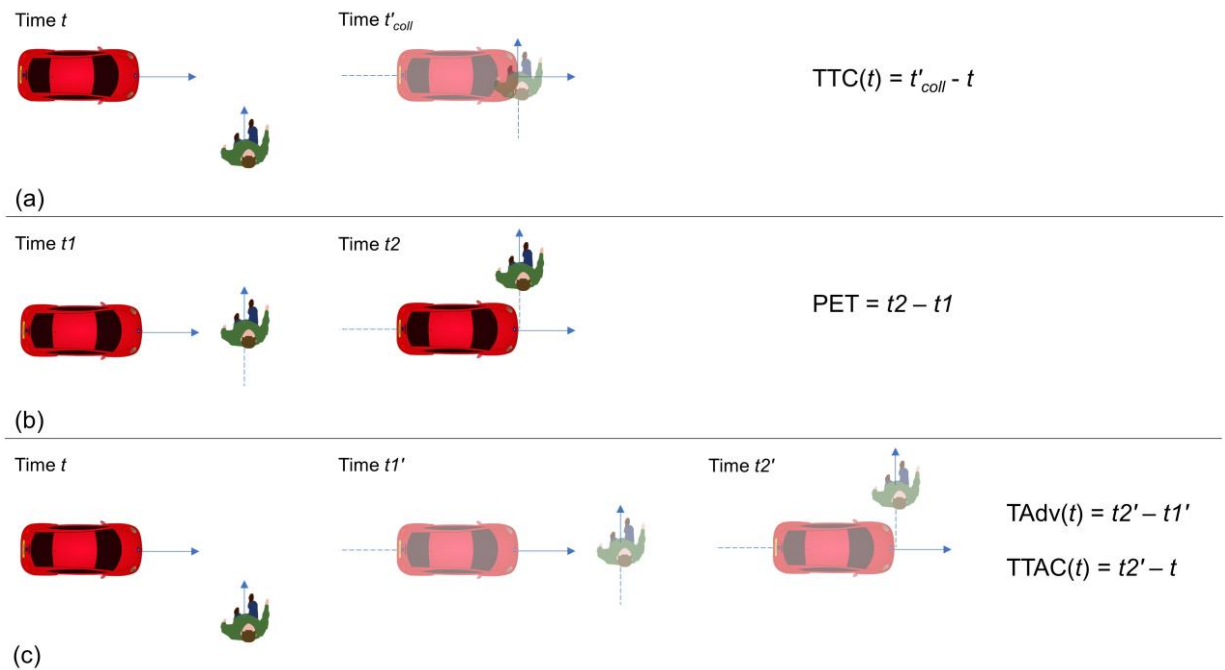

*Figure 1. Visual representation of: (a) time-to-collision (TTC), (b) post-encroachment time (PET), (c) time advantage (TAdv) and time-to-avoided-collision-point (TTAC). Solid figures identify observed situations, transparent figures identify predicted situations.*

## 2.3 From traffic conflicts to crash probabilities

Since early studies in traffic conflict theory, researchers have tried to connect the frequency of conflicts with that of crashes. In the earliest works, a single and fixed conflict ratio was sought (Hauer, 1982; Hydén, 1987). The idea was to compare crash and conflict data collected at various locations and to identify a single coefficient to link them. However, due to the strong heterogeneity between the conditions in which conflicts (collected over short periods of time) and crashes (aggregated over multiple years) were observed, this led to obtaining mixed results and highlighted the need to adopt more advanced approaches. An evolution of this approach consisted in recognizing the need to account for this heterogeneity and, instead of a fixed ratio, looking for a function of such conditions (El-Basyouny and Sayed, 2013; Guo et al., 2010; Sayed and Zein, 1999). This function could be estimated with count models, including conflict rates among explanatory variables.

An alternative approach is based on a rather different premise, according to which conflicts and crashes are the same type of event, characterized in a continuous way by different levels of severity, similarly to what was originally proposed by Glauz & Migletz (1980) (see Figure 2). From this, it is possible to hypothesize a probability distribution for the collision nearness of traffic events and, from that distribution, to extract a crash probability. Campbell et al. (1996) and Songchitruksa &

Tarko (2006) proposed to use extreme value distributions to obtain the probability of a crash to happen, given the observed conflicts. The most interesting aspect of this approach is the possibility to estimate this probability without the use of crash data, which are instead used only for model validation. In the last few years, extreme value theory (EVT) has been applied in various road safety works (Alozi and Hussein, 2022; Arun et al., 2022; Cavadas et al., 2020; Farah and Azevedo, 2017; Fu et al., 2021, 2020; Orsini et al., 2020, 2019; Wang et al., 2018; Zheng et al., 2018a, 2014, 2019), with more or less complex and sophisticated procedures. It has also been applied in the specific case of before-after road safety analyses (Zheng et al., 2018b; Zheng and Sayed, 2019), which, however, did not focus on vulnerable road users.

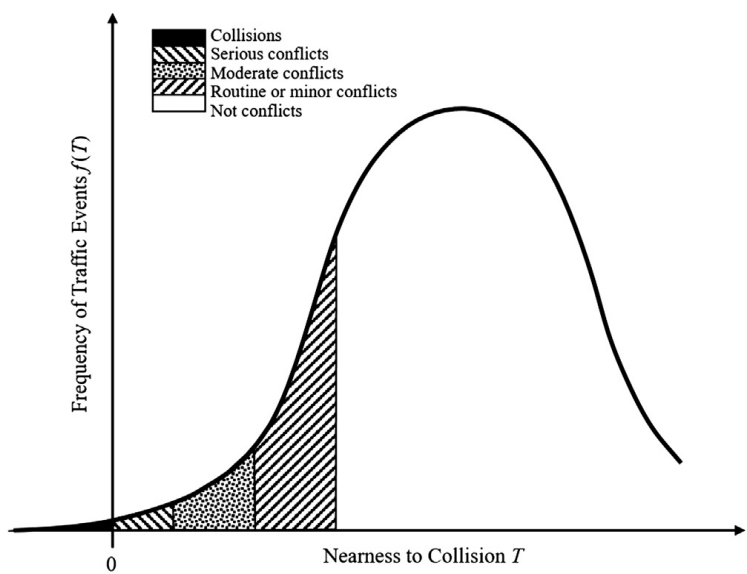

*Figure 2. Illustration of the concept of continuous distribution of crash nearness. Source: Tarko (2019), adapted from Glauz & Migletz (1980).*

Recently, Tarko (2018b) proposed the use of an extreme value distribution, called Lomax, which is essentially a particular type of Pareto distribution, shifted so that its domain begins at zero. The author compared different approaches for the estimation of parameters, with the use of synthetic data: Maximum Likelihood, Probability-Weighted Moments, and Single Parameter Estimation (SPE). The latter was found to be more accurate and efficient, and was applied in subsequent real-world case studies on road departures (Tarko, 2020) and rear-end collisions (Tarko and Lizarazo, 2021). For a thorough and rigorous theoretical justification of the distribution and the method applied, the reader is referred to Tarko (2018b, 2019, 2021); more details on the crash estimation procedure are given in Section 3.3.

# 3. Methodology

## 3.1 Methodological framework

The proposed methodology is illustrated in Figure 3, and aims to evaluate the safety of an "After Scenario", with reference to a "Before Scenario". The After Scenario is an evolution of the Before Scenario, in which an infrastructural (*ex post* evaluation) or a virtual (*ex ante* evaluation) intervention is carried out. In the present work, the focus is on the application of this methodology to before-after safety evaluation of shared spaces, but it can be generalized to a wide range of urban design interventions aimed at improving the safety of VRUs.

In the Before Scenario, data is collected with on-field surveys, while, in the After Scenario, with either on-field surveys (if the physical intervention has been already implemented) or simulation tools (if the physical intervention is yet to be implemented). From these surveys/simulations, vehicles and VRUs trajectories are extracted and archived into a trajectory database, which is input for both the spatial behavior analysis and the conflict analysis.

Spatial behavior analysis focuses on investigating changes in traffic volumes, road user trajectories, speed profiles, yield behaviors, and interaction locations. It is crucial for understanding how the new scenario influences the behavior of road users, as that could reliably suggest whether there is a general improvement in safety or highlight issues/unexpected consequences of the evaluated intervention.

To properly quantify the change in safety with a comprehensive indicator, a conflict-based analysis is carried out; to do so, conflicts are here treated in a quantitative way, in contrast to the qualitative approach of defining them based on observed evasive maneuvers. Therefore, an appropriate surrogate measure of safety must be selected. Section 2.2 discusses some measures for vehicle-VRU interactions and indicates time-to-the-avoided-collision-point (TTAC) as a particularly promising candidate. This choice was made with practical applications in mind, where measures relatively easy to compute, yet reliable, need to be used. It should be noted that more advanced approaches exist in the literature, for example combining information from several surrogate measures of safety (Ezzati Amini et al., 2022). More details on TTAC calculation are given in Section 3.2.

Traffic conflicts are then defined based on a fixed threshold of the selected surrogate measure of safety and used to obtain a crash risk estimate. The procedure is described in Section 3.3, and it is based on the assumption of a probabilistic relationship between conflicts and crashes.

This indicator is able to univocally determine whether the After Scenario is safer than the Before Scenario, and to quantify this change. If the result is negative, then the spatial behavior analysis results are essential for explaining it and suggesting alternatives to modify, when possible, the After Scenario. If the result is positive, then the spatial behavior analysis is still quite important to highlight the advantages and limitations of the new scenario, which could be useful for future works. In the case of prospective interventions evaluated with simulation tools, several After Scenarios can be developed and compared.

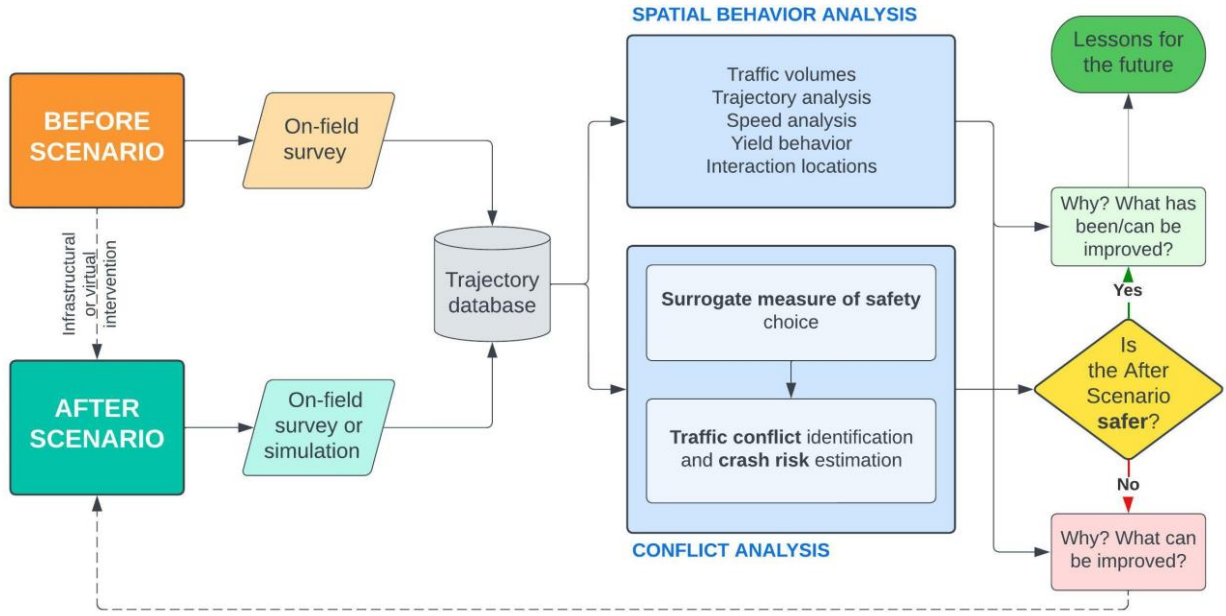

*Figure 3. Overall methodology framework*

## 3.2 TTAC calculation

Given the trajectories of two road users interacting with each other, discretized with a sufficiently high data collection frequency, we first consider a time instant *t* at the beginning of their encounter. Assuming constant direction and speed, the two predicted trajectories (which are linear segments) intersect at a conflict point CP(*t*) (Figure 4). Each road user needs some time to reach that point at their current speed; the first to do so is road user 1 at time *t1'(t)=d1(t)/v1(t)*, where *d1(t)* is the linear distance between road user 1 and CP(*t*), and *v1(t)* its speed. Similarly, the second road user will reach CP(*t*) at *t2'(t)=d2(t)/v2(t)*. The difference between *t2'(t)* and *t1'(t)* is TAdv(t) (see Section 2.2), and if it is equal to 0, the two users are on a collision course. TTAC(*t*) corresponds to *t2'(t)*, i.e. the time required for the second user to reach the conflict point; for this reason, Laureshyn et al. (2010) simply referred to it as "$T_2$". TTAC(*t*) is not computed if the first road user has already passed CP(*t*), as there is no collision risk involved anymore.

TTAC is then computed for each successive valid time instant; the severity of the interaction between two road user is given by the minimum value of TTAC computed during their encounter.

This calculation becomes more challenging when taking into account the road users' dimensions (more details on this are presented in Laureshyn et al., 2010). In the present work, TTAC was computed considering the road users as two moving points, i.e., the centroid of the VRU and each corner of the motor vehicle; therefore, for each vehicle-VRU encounter, four sets of TTAC values were obtained (i.e., one for each vehicle corner), and the lowest overall TTAC value was considered to quantify the severity of the interaction.

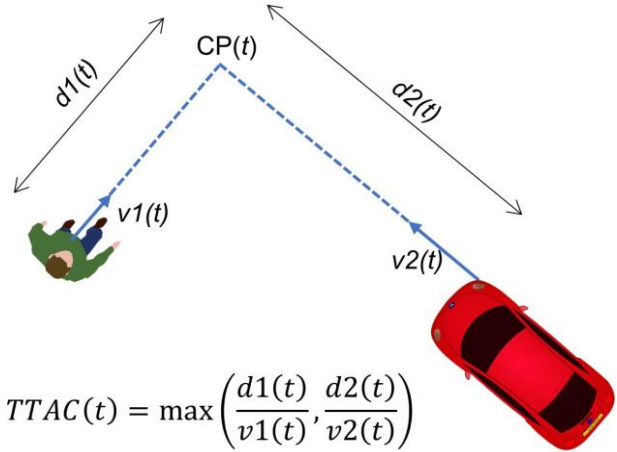

$$TTAC(t) = \max\left(\frac{d1(t)}{v1(t)}, \frac{d2(t)}{v2(t)}\right)$$

*Figure 4. Illustration of TTAC calculation*

## 3.3 Estimation of crash probabilities with Lomax distribution

The procedure to estimate crash probabilities applied in the present work is rather straightforward and could be realistically applied by practitioners in the real world. The use of SPE for the estimation of a Lomax distribution has the advantage of producing usable estimates with relatively fewer conflicts compared to other traditional EVT estimation methods, which tend to be affected by convergence issues (Tarko, 2018b). The procedure can be divided into four main parts.

1) Traffic conflicts are identified using multiple thresholds $u$ of the chosen surrogate measure of safety (e.g., TTAC). For each $u$, the total number of conflicts is $n$. The generic traffic conflict, defined by a TTAC value below the threshold, is called $c_i$ ($i = 1, …, n$). Exceedances $x_i$ are calculated as:

$$x_i = u - c_i \tag{1}$$

2) For each threshold value, a Lomax distribution is fitted with SPE. According to this method, the parameter $k$ of the distribution can be estimated as:

$$k = \frac{\sum_{i=1}^{n} \log\left(1 - \frac{i - 0.5}{n}\right) \log\left(1 + \frac{x_i}{u}\right)}{\sum_{i=1}^{n} \left[\log\left(1 + \frac{x_i}{u}\right)\right]^2} \tag{2}$$

3) Given the fitted distributions, for each threshold value, it is possible to estimate the conditional crash probability P, which corresponds to the cumulative probability of having $c_i = u - x_i \leq 0$, i.e. a crash, conditional to $c_i < u$, i.e., $c_i$ being a conflict:

$$P(c_i \leq 0 | c_i < u) = 2^{-k} \tag{3}$$

From (3), given n, the expected number of incidents $Inc_t$ in the observation period t is given by:

$$Inc_t = n * 2^{-k} \tag{4}$$

This could be extended to a longer time horizon T assuming constant collision probability throughout T:

$$Inc_T = \frac{T}{t} n * 2^{-k} \tag{5}$$

4) The final threshold u is chosen among all candidates. Here, it must be noted that in the real world a clear-cut threshold separating traffic conflicts from controlled interactions does not exist: choosing a high enough threshold would allow no conflict to be missed, but would also wrongly include many controlled interactions; conversely, choosing a low enough threshold to exclude all controlled interactions would also exclude some real conflicts. It can be demonstrated that the latter choice should be made to identify u, which means that the threshold u should be lower than the population minimum preferred crash nearness. In other words, it is equivalent to assuming that any observed encounter with a minimum TTAC lower than u is bound to be affected by some kind of failure (i.e., a temporary lack of response), as users would not intentionally accept that level of nearness. More details on this are given in Tarko (2021).

The effect of threshold choice can be visualized in Figure 5. Obviously, as the threshold decreases, the criterion for defining conflicts becomes stricter and the number of conflicts n decreases (Figure 5a); reducing the threshold rightly increases the conditional probability P, as more severe conflicts are more likely to result in a crash, but conversely too high threshold values produce biased results that inflate P (Figure 5b), due to the inclusion of controlled interactions in the fitting dataset; considering the resulting estimated number of crashes $Inc_t$, it is possible to observe a "stable" zone in which, reducing the threshold value, the corresponding increase in P is compensated by a decrease in n (Figure 5c). Choosing any threshold u in this zone would produce a similar estimated number of crashes. Normally the highest value of u within this zone is chosen (which roughly corresponds to the lower value of P), as it allows the most efficient use of data and consequently the most reliable estimation of the Lomax distribution parameter.

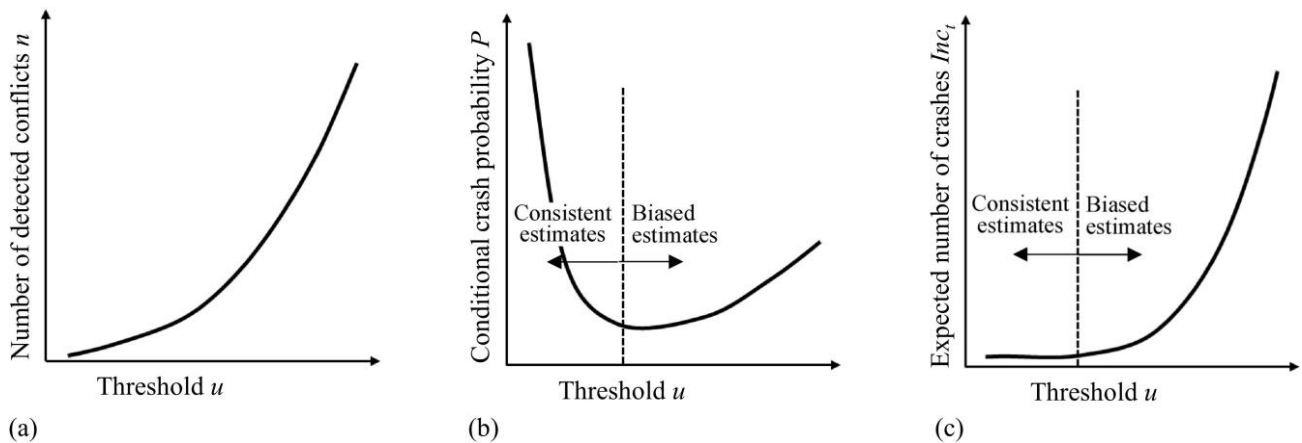

*Figure 5. Effect of threshold u on: (a) number of detected conflicts, (b) conditional crash probability, expected number of crashes. Adapted from* (Tarko, 2019).

# 4. Data preparation

## 4.1 Before and after scenarios

The study area is located in a residential zone of Sambruson di Dolo, a village in the Veneto region (Italy), with about 5,000 inhabitants. Figure 6a shows the Before Scenario; in the study area, there were three main attractors: a kindergarten, a bar and a cemetery, all served by a bidirectional road with a single sidewalk which also included a sharp 90° curve. The road served as the south branch of a roundabout, acting as a minor gateway toward the central part of the village, which is located in the north/east, with respect to the satellite view of Figure 6.

Figure 6b presents the area after the implementation of a series of urban design solutions. The main intervention was a shared space that covered most of the east side of the study area, including the 90° curve. A new access to the kindergarten was provided, linking it to a new bicycle lane and, thanks to a new raised pedestrian crossing, to the cemetery parking area.

Within the implemented shared space, all sidewalks and road markings were removed; the street pavement was remodeled, replacing the asphalt surface with red and grey concrete tiles. New elements of urban furniture such as benches and flower boxes were placed on the street, with the double effect of providing a partially protected "safe zone" to the VRUs and limiting vehicle passage to one per direction on the cross-section in front of the main access to the kindergarten (S3 in Figure 7 and Figure 8). A small flashing light, a roadway narrowing with short vertical poles, and traffic signs posting a 20 km/h speed limit and indicating right-of-way for pedestrians, were placed at both shared space entrances.

## 4.2 Data collection and processing

The Before Scenario survey was carried out using six video cameras on a sunny day in the fall of 2017 (Wednesday October 11th). The recordings were made during the morning (07:30 - 09:30) and the afternoon peak hours (15:00-16:45), which respectively contained the opening and closing time of the kindergarten. The After Scenario survey was carried out on a similar sunny day in the fall of 2020 (Wednesday November 18th), recording data with the same cameras in the same time slots.

Figure 7 presents some examples of the camera fields of view. The cameras were mounted on telescopic poles hooked to light poles, about eight meters above the ground. They were positioned, when possible, at the exact spot in both surveys; although some light poles used to support the cameras in the first survey were removed, the field of view of the videos was controlled to cover the same areas in both surveys.

Six notable cross-sections were identified; each camera was positioned to frame the vicinity of at least one cross-section (Figure 8). The video surveys covered the entire study area, collecting data both inside and just outside the shared space zone.

Video processing was performed using the methodology developed by Trifunović et al. (2021) and previously applied in other shared space studies (Batista and Friedrich, 2022a, 2022b). It is illustrated in Figure 9. Raw extracted trajectories were smoothed and transformed into 2D plane coordinates using a homography matrix. To achieve the required level of precision for a reliable conflict analysis, all trajectories involved in notable interactions (see Sections 5.1.4, 5.2) were manually inspected and, when necessary, corrected. The resulting trajectory database consisted of several tables (one for each survey and camera) containing the following fields: a trajectory ID, the video frame number, the X and Y coordinates (with respect to a 2D satellite picture), and a label indicating the road user type.

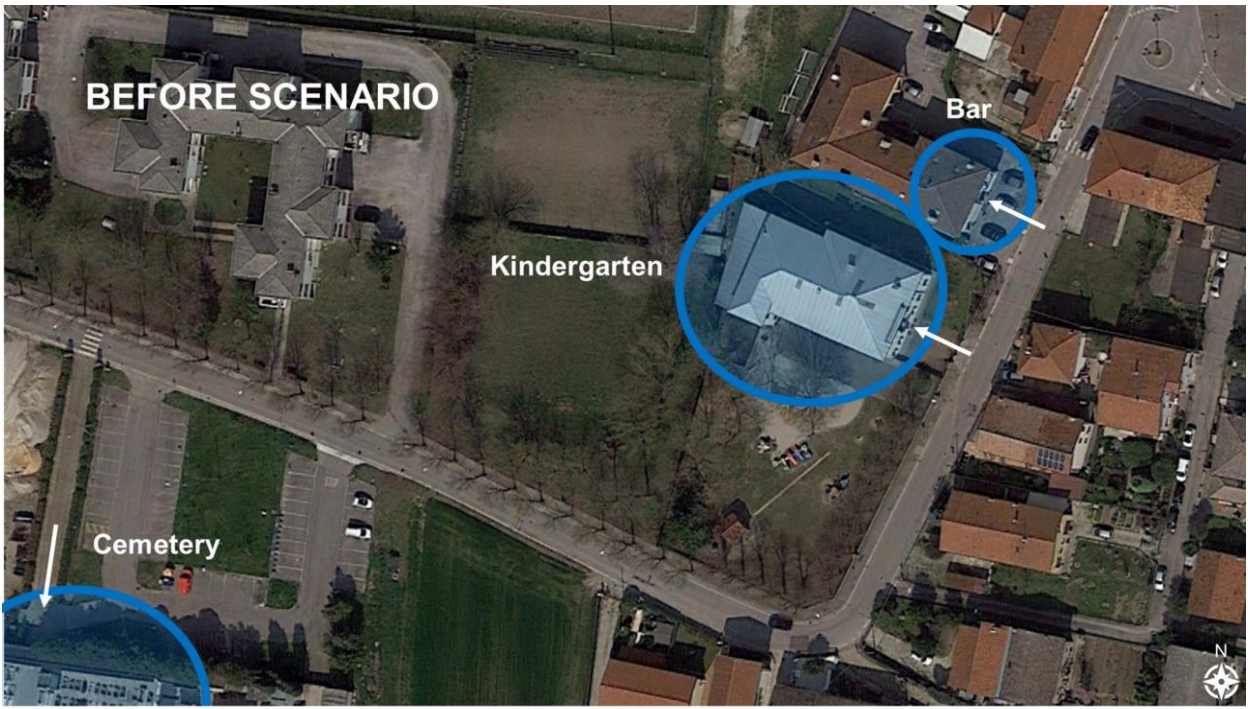
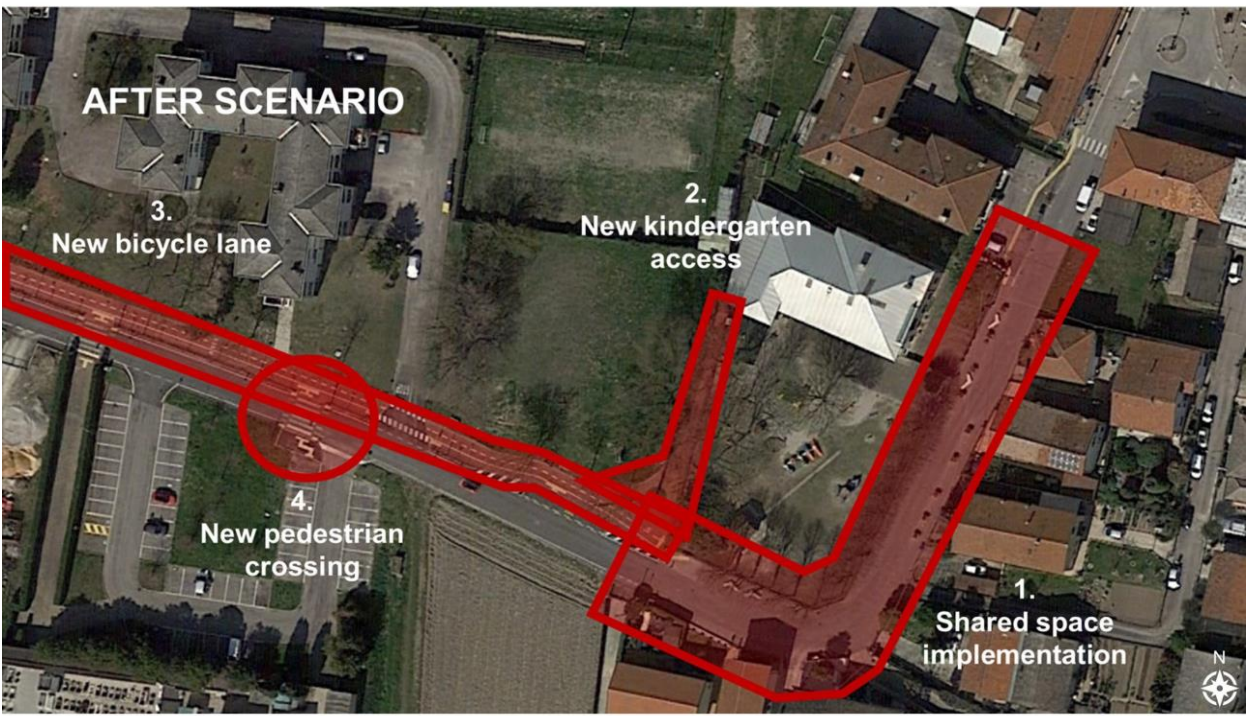

*Figure 6. Satellite view and description of (a): Before Scenario (white arrows identify access points), (b) After Scenario.*

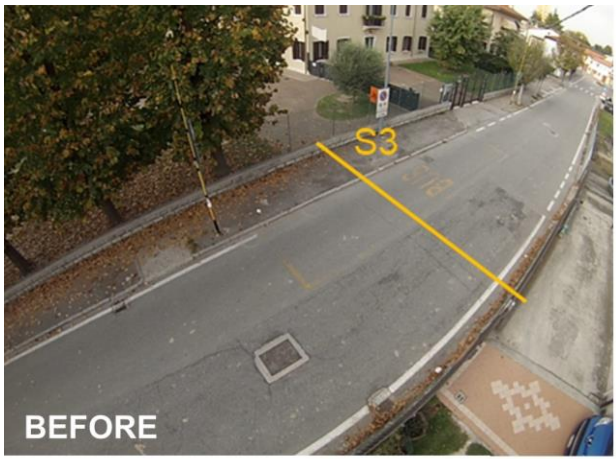 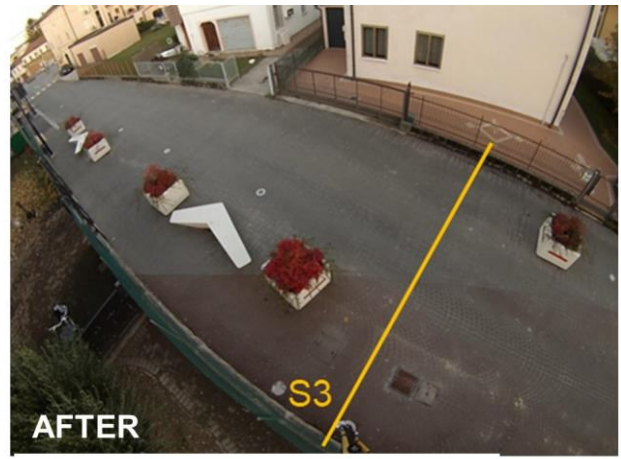
(a) (b)

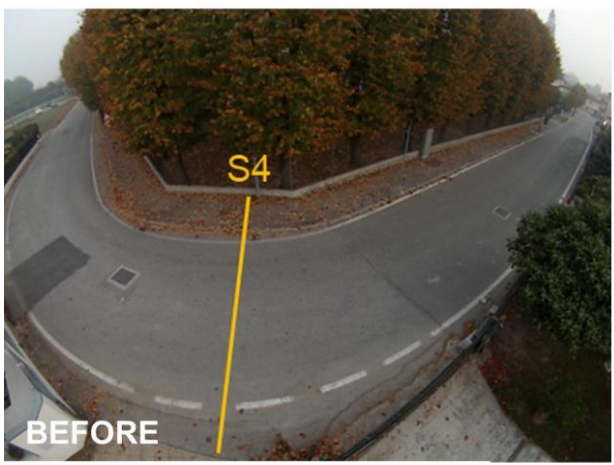 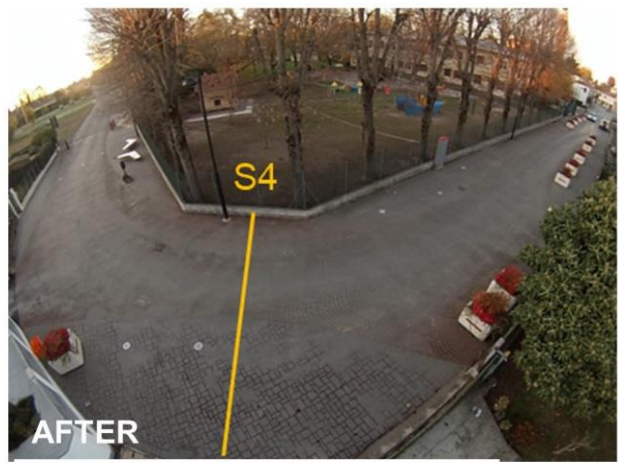
(c) (d)

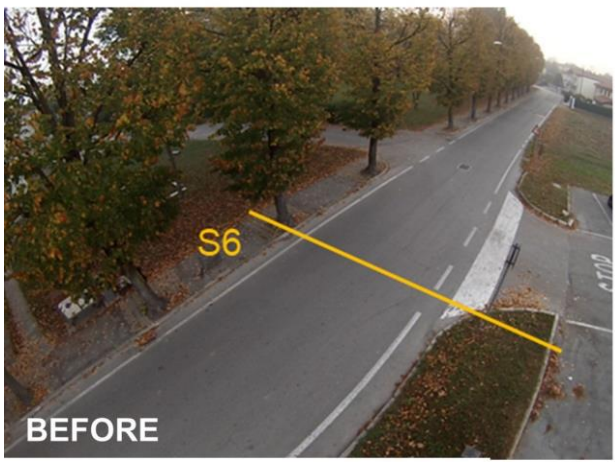 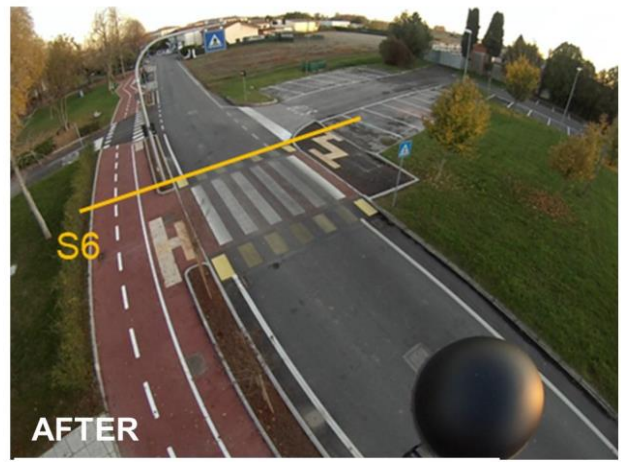
(e) (f)

*Figure 7. Before and After field-of-views of selected video cameras. Yellow lines and codes identify cross-section locations.*

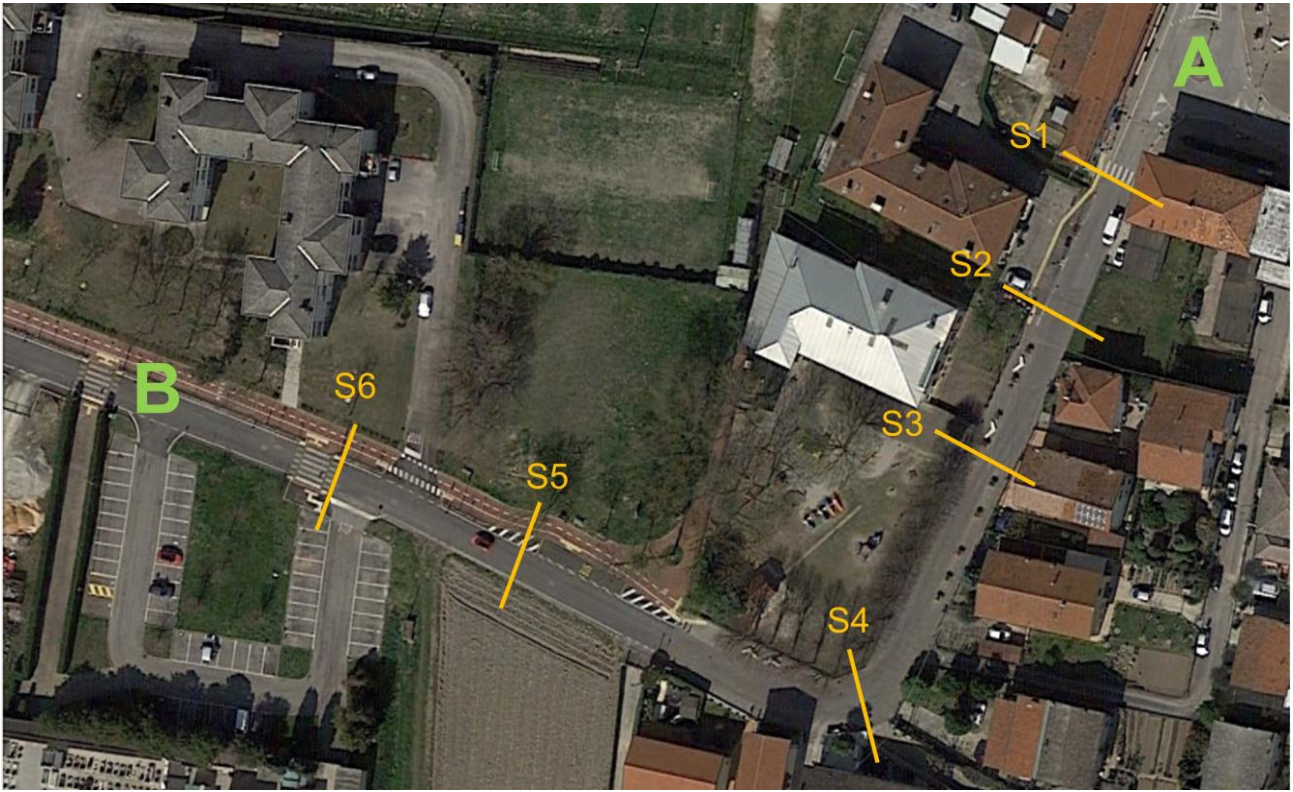

*Figure 8. Cross-section locations and codes.*

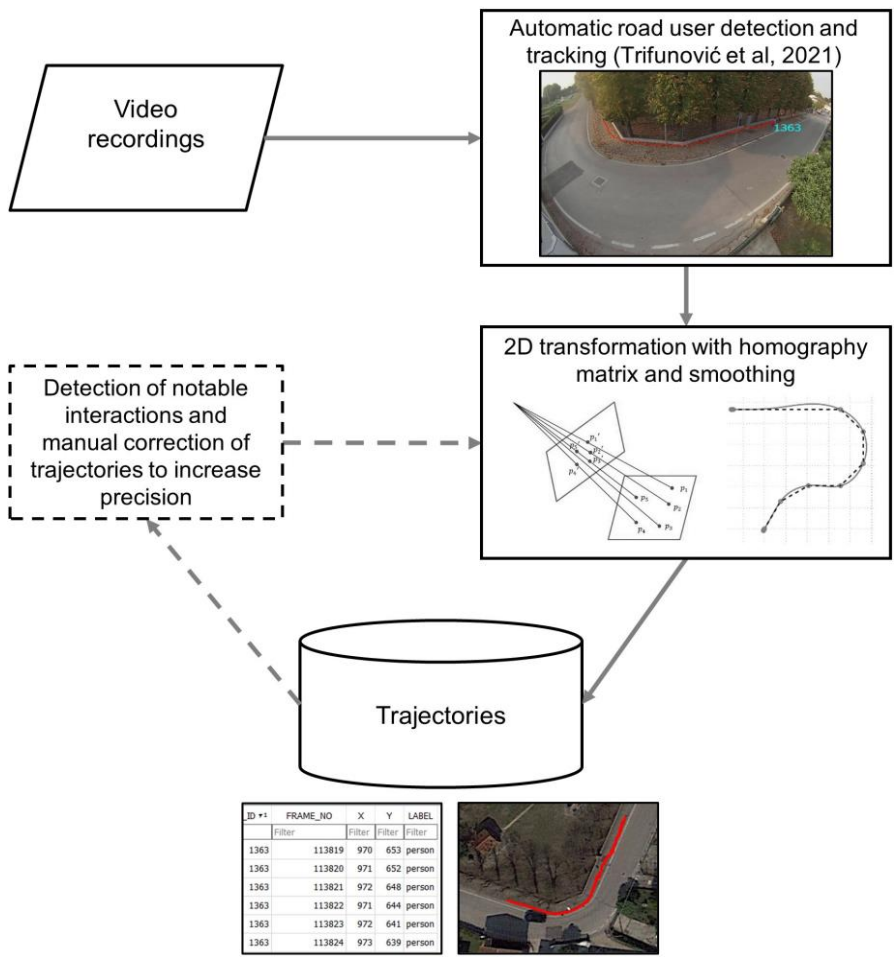

*Figure 9. Video data processing procedure. Solid lines represent the main procedural sequence, dashed lines the correction procedure, which was applied when necessary.*

# 5. Results

## 5.1 Spatial behavior analysis

### 5.1.1 Traffic volumes

Traffic volumes were analyzed at each cross-section in each scenario, considering both the direction from the central part of the village toward the west (from point A to point B, with respect to Figure 8) and the opposite direction (from B to A). Pedestrians and bicyclists were aggregated into a single class of vulnerable road users, due to the relatively small number of bicyclists observed (there was an overall ratio of about 5-to-1 between pedestrians and bicyclists).

Table 1 presents the volumes recorded in S3, which were representative also of S1, S2, and S4. There was an overall reduction of about 19% in motor vehicles, which was more apparent in the B→A direction (-25%). The number of VRUs crossing S3 was similar in both scenarios, yet a noticeable unbalance between the two directions (+38% A→B, -35% B→A) was observed.

Table 2 presents the volumes in S6, which are consistent with S5. S6 was positioned to detect all pedestrians' movement on the zebra crossings (Figure 7); trajectories intersecting the S6 cross-section multiple times were counted only once. Here again, it is possible to observe a decrease in vehicle volumes, similarly to S3. In contrast, the total number of VRUs almost quadrupled.

The main reason for the observed trends is likely imputable to the new access to the kindergarten and the new pedestrian crossing (Figure 6b). In the Before Scenario, many parents arriving from point B used to momentarily stop the car in front of the old kindergarten entrance and then continue on their path, whereas in the After Scenario it was more convenient to park at the cemetery, cross on foot the pedestrian crossing (i.e., S6), walk the children to the kindergarten, and then walk back to the car. The decrease in the A→B direction is less evident because parents arriving from point A still need to cross S3 and S6 to enter the parking lot.

The overall reduction in the volumes of motor vehicles in the study area can be imputed to multiple factors: (i) the effect of shared space, with people more willing to walk or choose alternative paths; (ii) the fact that the cemetery parking could be reached by other paths, bypassing the shared space; (iii) the effect of the COVID-19 pandemic, which is discussed in Section 6.2.

*Table 1. Traffic volumes at cross-section 3 in each scenario.*

| S3 Direction | Motor vehicles | | Vulnerable road users | |
|---|---|---|---|---|
| | Before Scenario | After Scenario | Before Scenario | After Scenario |
| A→B | 121.1 veh/h | 105.3 veh/h (-13%) | 32.8 vru/h | 45.3 vru/h (+38%) |
| B→A | 100.8 veh/h | 75.5 veh/h (-25%) | 43.5 vru/h | 28.3 vru/h (-35%) |
| Total | 221.9 veh/h | 180.8 veh/h (-19%) | 76.2 vru/h | 73.6 vru/h (-3.5%) |

*Table 2. Traffic volumes at cross-section 6 in each scenario.*

| S6 Direction | Motor vehicles | | Vulnerable road users | |
|---|---|---|---|---|
| | Before Scenario | After Scenario | Before Scenario | After Scenario |
| A→B | 124.1 veh/h | 112.8 veh/h (-9.4%) | 9.1 vru/h | 41.3 vru/h (+356%) |
| B→A | 80.5 veh/h | 62.1 veh/h (-23%) | 9.3 vru/h | 29.3 vru/h (+214%) |
| Total | 205.1 veh/h | 175.0 veh/h (-15%) | 18.4 vru/h | 70.7 vru/h (+284%) |

### 5.1.2 Pedestrian trajectories

A more detailed and disaggregated analysis was carried out on pedestrian trajectories to better understand the effect of the shared space implementation on their spatial behavior. Figure 10 shows how the trajectories changed in two cross-sections of interest.

Cross-section S3 was located in the middle of the shared space, in front of the old kindergarten entrance. In the Before Scenario (Figure 10a) pedestrians tended to stay and move almost exclusively on the available sidewalk; occasionally they used the roadway, especially in periods with high pedestrian density. In the After Scenario (Figure 10b), pedestrians still preferred to stay on the western side of the road, but the number of detected trajectories on the eastern edge increased, and, more importantly, also those walking along or crossing the roadway. This shows a significant effect of the implementation of the shared space. Yet, the installation of benches and plant boxes, which creates "safe zones" in some parts, preserved an element of road user segregation that goes against the concept of shared space.

In the Before Scenario, the few pedestrians crossing the road around S6 did so in an area without any pedestrian crossing facility (Figure 10c), which was rather dangerous, considering the speed of motor vehicles in this section (see Section 5.1.3). In contrast, in the After Scenario, with the implementation of a raised pedestrian crossing (Figure 10d), pedestrian volumes strongly increased, as discussed in Section 5.1.1; here the location of this provision was probably not ideal, as the trajectories show many pedestrians not complying with it, but some physical restrictions (e.g., the access road on the other side of the parking lot) also prevented alternative solutions.

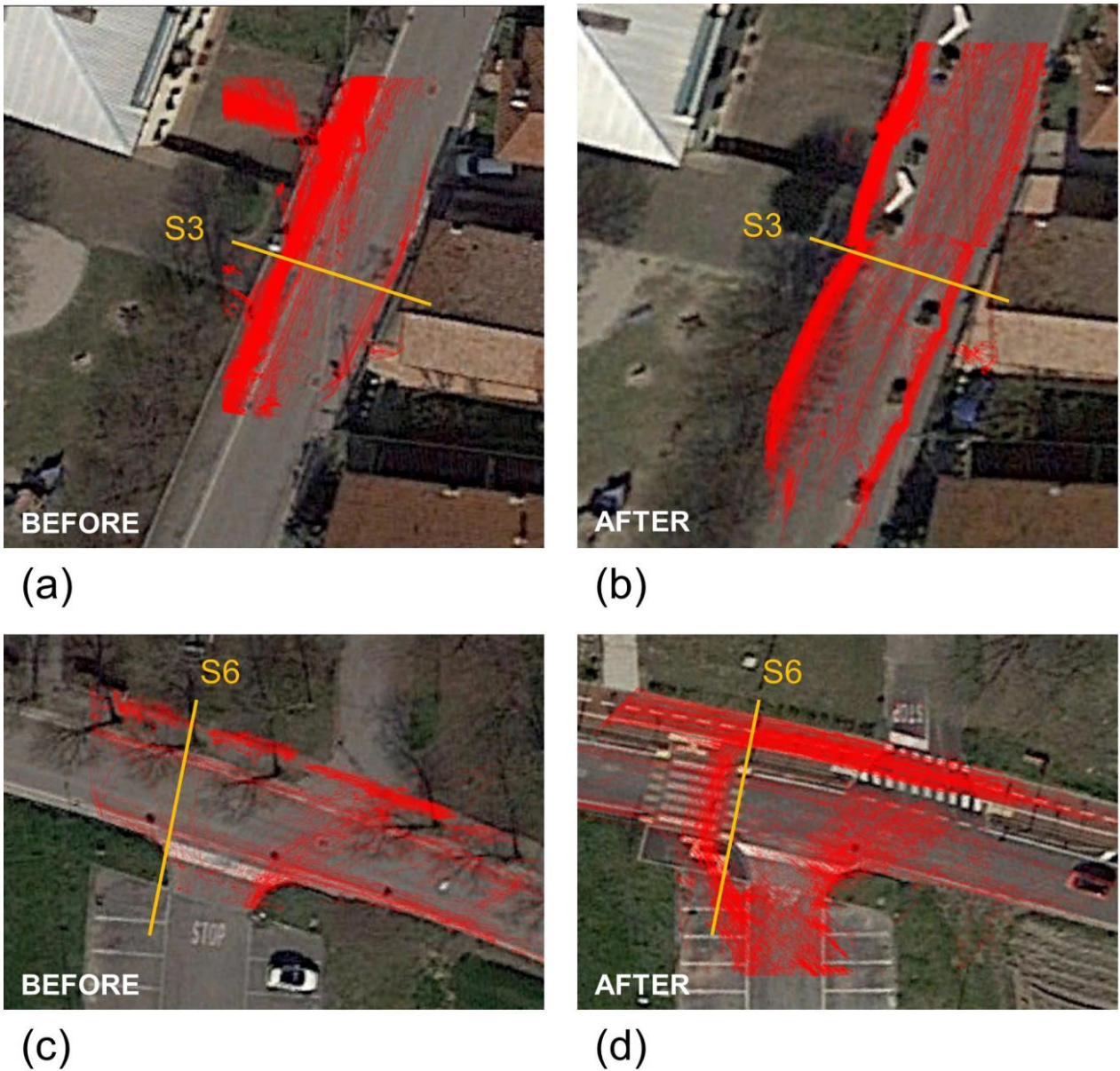

*Figure 10. Pedestrian trajectories at: (a) S3 in the Before Scenario, (b) S3 in the After Scenario, (c) S6 in the Before Scenario, (d) S6 in the After Scenario.*

### 5.1.3 Vehicle speed

One of the means toward an increase in VRU safety in shared spaces is the reduction in motor vehicle speed. Figure 11 presents the speed profiles in both directions, computed by averaging the speed detected at each cross-section. In the Before Scenario (Figure 11a) vehicles used to accelerate out of the roundabout (point A in Figure 8) until exceeding 30 km/h in front of the kindergarten entrance (S3), then performing a relatively strong deceleration due to the sharp 90° curve in S4, and, subsequently, accelerating to more than 40 km/h in S6. In the After Scenario, the effect of the shared space is immediately observable in S1, with vehicles exiting the roundabout at a 3 km/h lower speed, and keeping a constant speed of about 26 km/h (albeit above the prescribed 20 km/h) until exiting the shared space then accelerating above 30 km/h, before slowing down for the new pedestrian crossing.

A similar profile was observed in the opposite direction (Figure 11b). In particular, in the Before Scenario, vehicles strongly decelerated from 44.4 km/h in S5 to just 19.9 km/h in S4, whereas in the After Scenario, they kept a much more stable speed profile.

The non-parametric Mann-Whitney $U$ test were carried out to compare the speeds in each cross-section (Table 3), the null hypothesis being that the distributions of speed were identical in both scenarios. This confirmed a statistically significant ($p$-value < 0.05) lower speed in almost all cross sections in the After Scenario. The exceptions are S1 in the B→A direction, where the speed is likely bounded by the presence of the roundabout, and S4 in both directions, where the speed significantly increases in the After Scenario.

These results show in a very clear way the positive effects of the shared space and other urban design interventions (notably, the raised pedestrian crossing) on vehicle speed. In S6, for both directions, the speed almost halved, and this is especially relevant because a considerable number of pedestrians used to cross in that area in the Before Scenario; even though many pedestrians still did not comply with the pedestrian crossing in the After Scenario (Section 5.1.2), the speed reduction of motor vehicles caused by this provision is still positive in terms of safety. In S4, on the other hand, there was a significant increase in speed in the After Scenario, likely due to the widening of the roadway available to vehicles as a consequence of the removal of the sidewalk (Figure 7); this is potentially a safety concern.

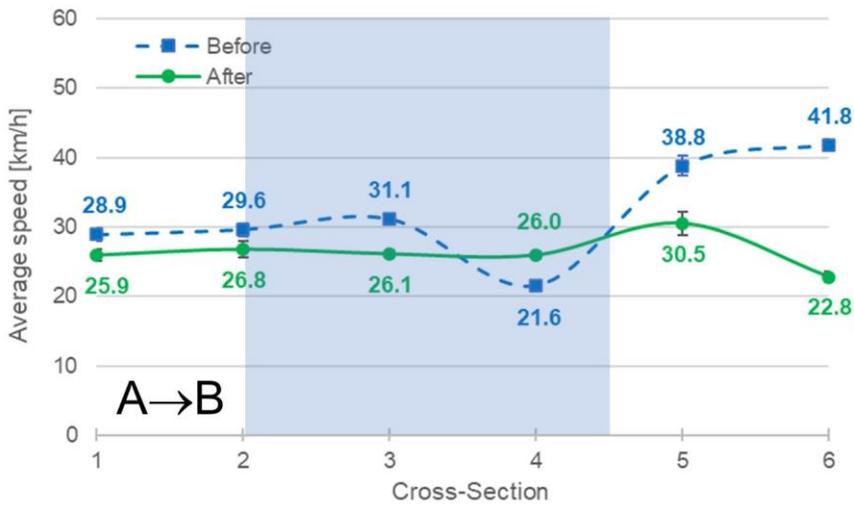

(a)

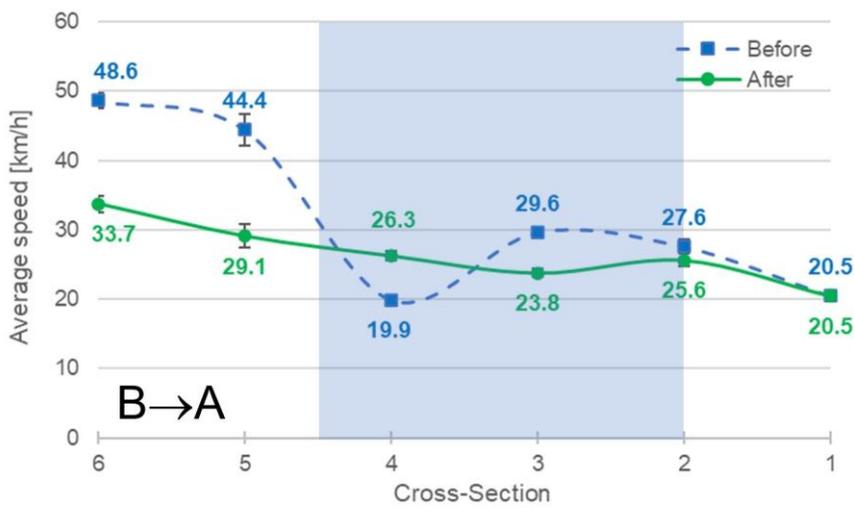

(b)

*Figure 11. Vehicle speed profiles in Before and After scenarios for (a) direction A→B, (b) direction B→A. Markers represent mean values, bars 95% confidence intervals of means. The shaded parts of the graphs identify the shared space*

*Table 3. Results of Mann-Whiney U test to analyze cross-section speed differences. Effect size is given by the rank biserial correlation (positive values: the speed in the Before Scenario tends to be higher than in the After Scenario)*

|               | A→B     |         |             | B→A     |         |             |
| Cross-section | statistic | p-value | effect size | statistic | p-value | effect size |
|---------------|---------|---------|-------------|---------|---------|-------------|
| S1            | 114,035 | <.001   | 0.185       | 70,723  | 0.698   | -0.016      |
| S2            | 91,996  | <.001   | 0.147       | 51,399  | 0.004   | 0.127       |
| S3            | 128,516 | <.001   | 0.437       | 80,588  | <.001   | 0.512       |
| S4            | 39,558  | <.001   | -0.567      | 17,353  | <.001   | -0.681      |
| S5            | 72,696  | <.001   | 0.336       | 38,642  | <.001   | 0.524       |
| S6            | 159,642 | <.001   | 0.863       | 60,195  | <.001   | 0.777       |

**5.1.4 Interaction locations**

As mentioned and motivated in Section 2.2, in this work, a surrogate measure of safety called TTAC is used to identify traffic conflicts. The selection of the TTAC threshold to separate traffic conflicts from controlled traffic interactions is part of the conflict analysis presented in Section 5.2. Here, the objective is to discuss the location of, more generally, "notable interactions". Given the qualitative nature of the present analysis, the TTAC threshold to identify these interactions was also chosen qualitatively, after evaluating a selection of video recordings; after this assessment, a TTAC threshold of 4 seconds was deemed appropriate, as it was high enough to include all vehicle-VRU encounters in which there was some kind of interaction between the users. It must be very clearly noted that, contrary to traffic conflicts, these interactions were not necessarily dangerous.

Figure 12 illustrates the location of notable interactions on the map; the figure also highlights proper conflicts (as defined according to Section 5.2), which at this stage should be treated, more generally, as notable interactions. In the Before Scenario (Figure 12a) there is a very dense concentration of interactions near the pedestrian crossing close to the roundabout, and others are more or less equally distributed on the road in front of the kindergarten; only a handful of interactions were detected on the location of the future crossing in S6, consistent with the low pedestrian volumes observed there (Figure 10, Table 2).

In the After Scenario (Figure 12b), the number of interactions at the roundabout crossing decreased significantly, which is consistent with the fact that a decrease in users speed, as observed in S1-S2 (Figure 11), tends to increase the TTAC values. The concentration of interactions in the shared space did not appear to change much, which is not necessarily a good thing, as more interactions are expected with a design approach encouraging user integration. It is, however, somewhat coherent with the remarks made in Section 5.1.2 that the position of benches/plant boxes maintains an element of segregation. The increase in interactions on the 90° curve may be linked to the increase in vehicle speed in that area (Figure 11) and could indicate some safety issues. On the other hand, an increase in interactions at the new pedestrian crossing is obviously expected.

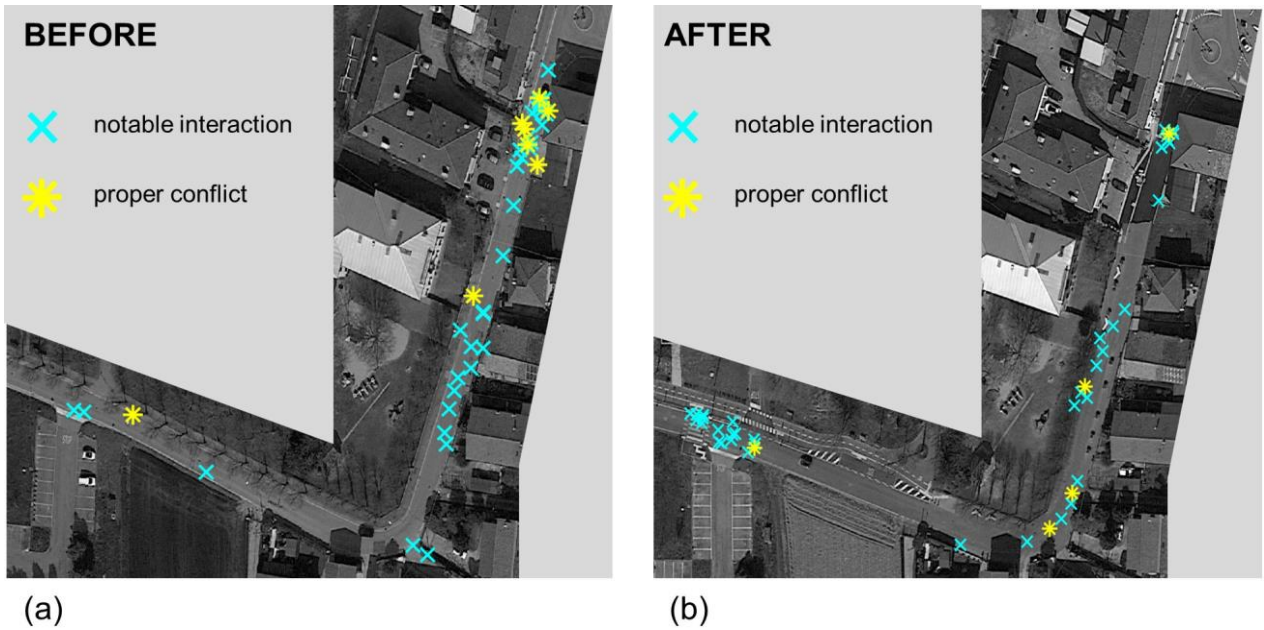

*Figure 12. Location of notable interactions (TTAC < 3 s) and proper conflicts (identified by TTAC thresholds defined in Section 5.2) in: (a) Before Scenario, (b) After Scenario.*

### 5.1.5 Yield behavior

A yield behavior analysis was carried out to assess the dominance of motor vehicles over VRUs. For this purpose, all vehicle-VRU encounters within the study area with PET less than 5 seconds (Navarro et al., 2022; Zangenehpour et al., 2016) were first considered. In the Before Scenario, in 94 out of 253 encounters (37,2%), the vehicle arrived at the encroachment point after the VRU, therefore giving right-of-way. The vehicle yield ratio increased in the After Scenario and in 104 out of 249 encounters (41.8%) the vehicle was detected at the encroachment point after the VRU; however, a statistical analysis performed with the Chi-squared test did not show any statistical significance. A more in-depth analysis, separating all encounters taking place in or out of the shared space, revealed that the change in yield behavior within the shared space was confirmed to be non-significant, with a slight negative trend. Out of it, the ratio increased from 40.9% to 52.6% [$\chi^2$ (1, $N$=213) = 2.73, $p$ = .09, marginally significant] near the roundabout, and from 15.6% to 40.8% [$\chi^2$ (1, $N$=108) = 6.42, $p$ = .01, significant] near the new pedestrian crossing. This confirmed the relevant impact of the new pedestrian crossing and suggested that the shared space was not particularly effective in reducing vehicle dominance within its boundaries, although having in fact an impact outside of it thanks to the reduction in vehicle speed (see also Sections 5.1.3 and 5.1.4).

A more advanced yield behavior analysis was carried out considering all interactions detected using a different surrogate measure of safety (TTAC, see Sections 2.2, 3.2, 5.1.4). This provided the opportunity to only focus on interactions that could potentially (but not necessarily) have been dangerous. Contrary to PET, which tells what happened but gives no indications on what could have happened, TTAC helps identify in which order the two users would have passed on the encroachment point if their movement remained unchanged. In this way, it is possible to compare the forecasted encroachment order at the time instant with the minimum TTAC, with the order effectively observed at the encroachment. Contingency tables were built for each scenario (Table 4, Table 5), considering all interactions with TTAC < 4 seconds in which the two trajectories eventually encroached.

In the Before Scenario (Table 4), 20 vehicle-VRU encroachments were forecasted and 17 were observed, with effectively 3 vehicles slowing down to give right-of-way to the VRU. A statistical analysis using the McNemar test, the Chi-squared equivalent for paired binomial values (McNemar, 1947), was then performed. Since the number of values in the discordant cells was lower than 25, the mid-*p* version of the test was used (Fagerland et al., 2013; Pembury Smith and Ruxton, 2020). According to this test, the change in the ratio between the two types of encroachment was non-significant ($p = .13$). In contrast, in the After Scenario (Table 5), 24 vehicle-VRU encroachments were forecasted and 18 observed, with 7 vehicles giving right-of-way to VRUs, and one VRU yielding to a vehicle; in this case, the change was statistically significant ($p = .04$).

This analysis showed that, by focusing the yield behavior analysis only on notable interactions, the After Scenario was safer, as the observed ratio of vehicles encroaching the avoided collision point after the VRU was significantly higher than the forecasted ratio.

*Table 4. Contingency table between observed and forecasted encroachments in the Before Scenario.*

| *Before Scenario* | | Observed encroachment | | TOT |
|---|---|---|---|---|
| | | vehicle-VRU | VRU-vehicle | |
| Forecasted encroachment | vehicle-VRU | 17 | 3 | 20 |
| | VRU-vehicle | 0 | 13 | 13 |
| | TOT | 17 | 16 | 33 |

*Table 5. Contingency table between observed and forecasted encroachments in the After Scenario.*

| *After Scenario* | | Observed encroachment | | TOT |
|---|---|---|---|---|
| | | vehicle-VRU | VRU-vehicle | |
| Forecasted encroachment | vehicle-VRU | 17 | 7 | 24 |
| | VRU-vehicle | 1 | 10 | 11 |
| | TOT | 18 | 17 | 35 |

## 5.2 Conflict analysis

As spatial behavior analysis was crucial for understanding the effects of shared space implementation on road users, it could not univocally determine whether there was a tangible improvement in road safety. Here, the conflict-based approach described in Section 3.3 was applied, with the objective of estimating the expected number of crashes in the Before and After scenarios, as a single comprehensive indicator of safety performance.

The procedure was carried out with a MATLAB script written by the authors. For each scenario, several Lomax distributions were fitted with the SPE method, starting with a proposed TTAC threshold of 3 seconds and then progressively reducing it with 0.05-second decrements. The effect of the threshold on the number of detected conflicts, the conditional crash probability, and the expected number of crashes is shown in Figure 13 and Figure 14. The observed trends in both scenarios were consistent with the expected theoretical trends in Figure 5, implying that the method chosen was appropriate given the distribution of the TTAC values detected in this case study.

In the Before Scenario (Figure 13), $P$ steadily decreased up to a 1.60-second threshold, indicating that a higher threshold would have included interactions that were not in fact conflicts, inflating the estimated number of crashes. Conversely, for lower threshold values, the expected number of crashes $Inc_t$ remained constant, around a 0.004 value. For $u = 1.60$ s, the estimated parameter $k$ was 10.9, $n = 8$, $P = 0.00054$, and $Inc_t = 0.0043$. This value represents the expected number of crashes in the observation period; to understand whether this result is plausible, the annual crash rate was extracted by multiplying $Inc_t$ by 200 (the number of days of the year during which the kindergarten is open). Thus, according to the model, there were about 0.86 expected crashes per year during school days peak hours. Of course, this value is far from being reliable, due to the very small and homogeneous observation period. For a more trustworthy annual rate, more observations should be carried out, in different weather conditions and seasons, and including off-peak periods; however, it is important for the credibility of this before-after analysis that the value obtained is plausible.

In the After Scenario (Figure 14), the trends of $P$ and $Inc_t$ identified a TTAC threshold of 1.70 seconds, which returned an estimated parameter $k$ of 11.3, $n = 5$, $P = 0.00039$, and $Inc_t = 0.0020$, i.e., 0.40 expected crashes per year during school days peak hours, indicating a -53% decrease from the Before Scenario, and relevant overall safety improvement in the study area.

The fact that the TTAC threshold is lower in the Before Scenario, indicates that road users accepted to interact with each other in a more "aggressive way" than in the After Scenario, i.e., that they had higher risk propensity and/or lower risk perception (Tarko, 2020). It should be noted that this does not mean that in the After Scenario the users interact less or that they are more spatially separated, as the TTAC value depends not only on spatial separation but also on road users' speed. The increase in the TTAC threshold in the After Scenario is consistent with a decrease in crash risk, as observed in other Lomax distribution applications in road safety (Tarko, 2020; Tarko and Lizarazo, 2021).

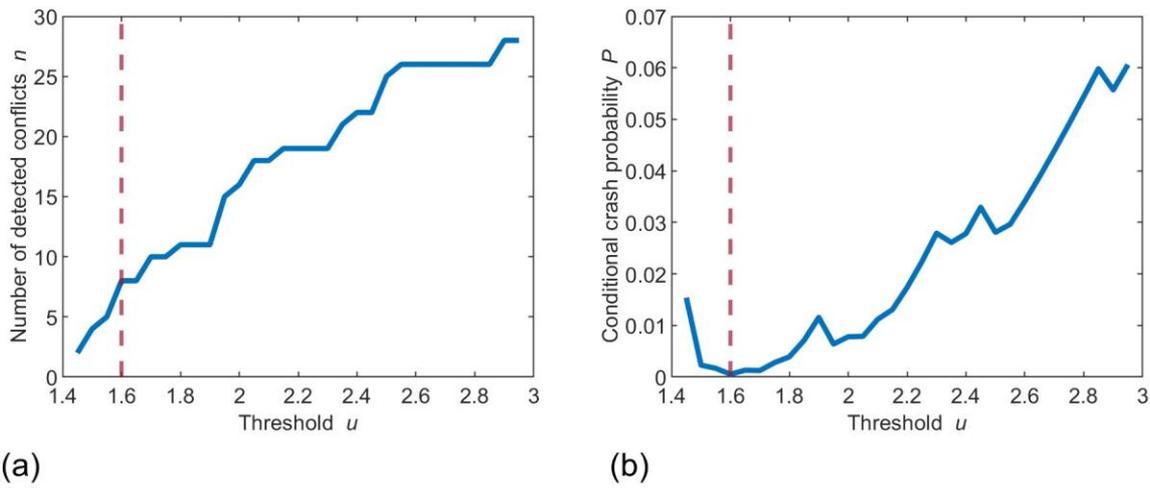

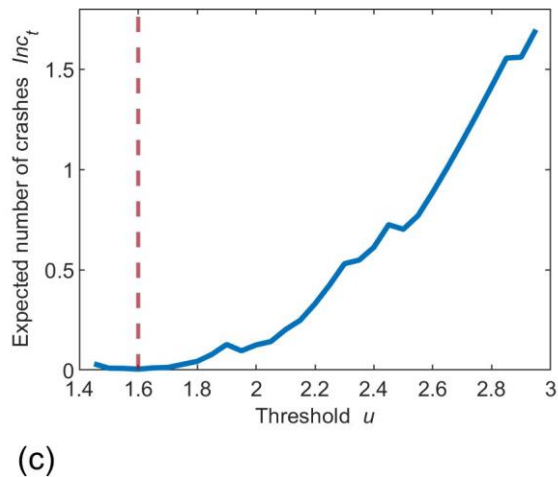

*Figure 13. Conflict analysis of the Before Scenario. Effect of TTAC threshold selection on: (a) number of detected conflicts, (b) conditional crash probability, (c) expected number of crashes.*

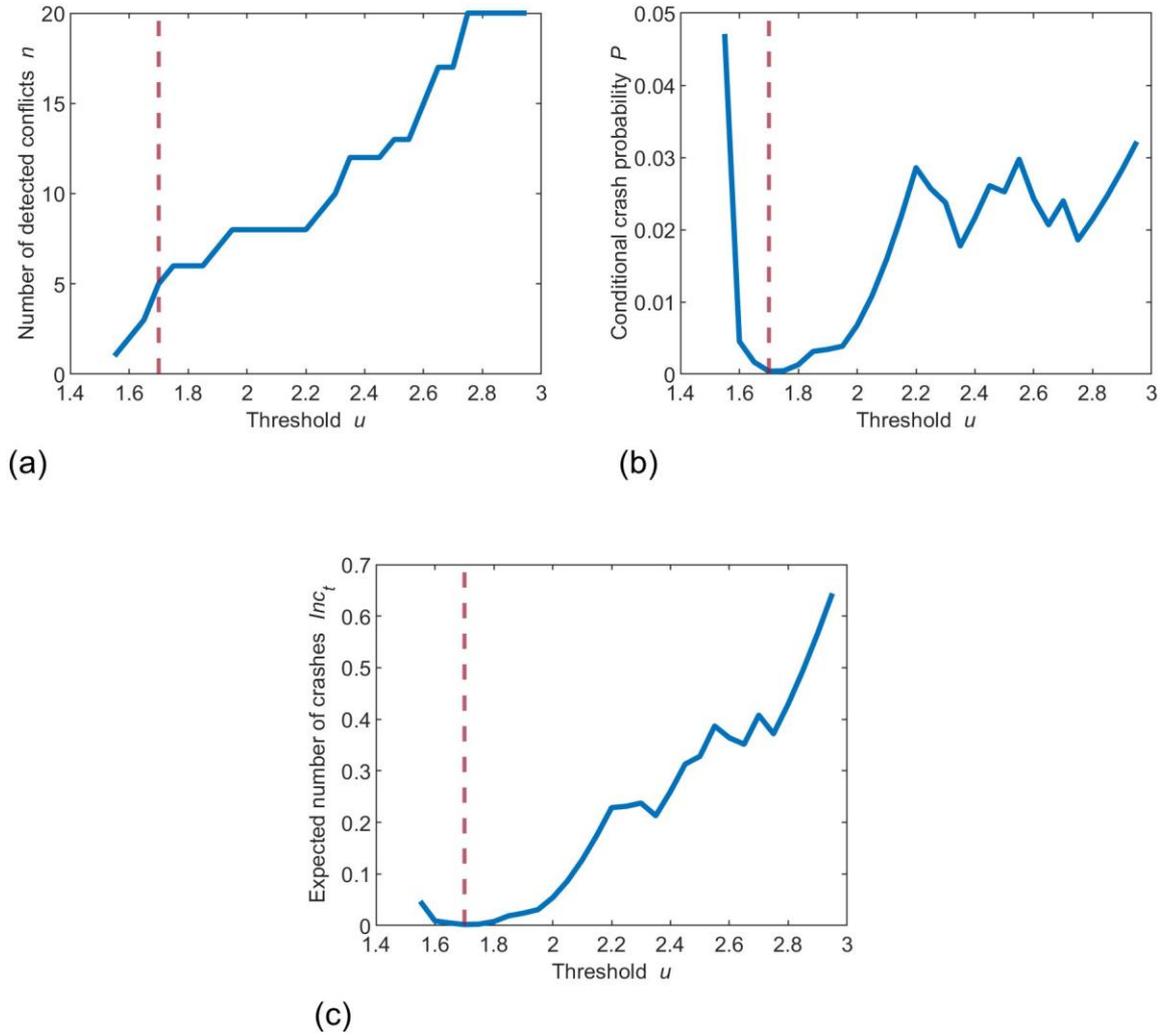

*Figure 14. Conflict analysis of the After Scenario. Effect of TTAC threshold selection on: (a) number of detected conflicts, (b) conditional crash probability, (c) expected number of crashes.*

# 6. Discussion

## 6.1 Main features of the proposed methodology

The application of the Methodology proposed in Section 3 was illustrated with a real-world case study in Sections 4 and 5, which highlighted some of its main features.

First, a single comprehensive indicator allowed to univocally determine the improvement in road safety. The spatial behavior analysis provided an understanding of the effects of the urban design interventions on driving behavior, indicating both positive and negative consequences. On the one hand, in the After Scenario, pedestrians tended to rather make broader use of the roadway area, the vehicle speeds were mostly reduced, fewer interactions were observed in the pedestrian crossing at the roundabout, and the yield behavior improved; on the other hand, many pedestrians crossed the road outside the new pedestrian crossing, some segregation between road users remained, vehicle speed increased at the 90° curve, and the improvements in yield behavior appeared to be relevant only in the areas just outside the shared space. In this case, spatial behavior analysis alone would not have been able to provide a clear and definitive indication of road safety. Besides, even if all indicators would have been pointing toward a road safety improvement, it would not have been possible to quantify this improvement; this would be a serious limitation in the case of a hypothetical comparison of several virtual After Scenarios.

Second, the combination of conflict and spatial behavior analyses is fundamental for an exhaustive assessment of the shared space implementation. Conflict analysis alone cannot provide additional information other than the expected number of crashes. In the present case study, the spatial behavior analysis highlighted some potential issues to address which, once solved, could even contribute to further improving safety conditions.

Third, the proposed methodology is relatively simple and could realistically be applied by practitioners in real-world projects. In road safety literature, more advanced before-after methodologies have been proposed, for example accounting for unobserved heterogeneity (Tahir et al., 2022) but, for realistic transferability from research to practical applications, the methodology must be compatible with the limited time and resources available for this kind of urban design interventions which, as in the present case study, have relatively low costs. The proposed methodology is based on theoretically-sound choices on the surrogate measure of safety to identify traffic conflicts and on the distribution used to model the probabilistic relationship between conflicts and crashes (as extensively discussed in Section 2), but it is also easily applicable: e.g., given the user trajectories, the computation of the TTAC is almost trivial, the SPE for the Lomax distribution consists in applying a single formula, and the threshold choice procedure is quite intuitive. The main practical obstacle may be the extraction of raw trajectories from the video recording, which requires some experience from the analyst and some time-consuming manual corrections, but this may be improved in the future considering the progress in developing tools to automatically extract trajectories, and should not be a problem for the evaluation of scenarios with the use of simulation tools.

Finally, the methodology is quite flexible. As shown in the present case study, it can be used to evaluate the safety of shared spaces, but it can also be applied to any scenario where a before-after evaluation of VRU safety is needed. The spatial behavior analysis might then include additional elements (e.g., investigation of pedestrian speeds, vehicle decelerations, etc.). Even if the choice of TTAC indicator and Lomax distribution seems to be particularly appropriate for this kind of application, the surrogate measure of safety used to identify conflicts does not necessarily need to be TTAC. Other more or less complex indicators may be used, and the applied EVT approach could also be different and possibly more advanced.

## 6.2 Case study limitations

The case study presented in this work is affected by some limitations which must be pointed out, but that can also serve as useful takeaway lessons for practical applications (Section 6.3).

The main limitation was the relatively small number of observed conflicts, which had several consequences. Even though the crash estimations of the present case study were plausible (as discussed in Section 5.2), the reliability of Lomax estimates obtained with such a small fitting dataset could be questionable, since a single very severe conflict could potentially inflate the results. In fairness, it should be pointed out that, differently from other EVT approaches that usually require, as a rule of thumb, at least 30 observations in the fitting dataset (Farah and Azevedo, 2017), the Lomax distribution has been shown to provide usable and reliable results even with 20 observations (Tarko, 2018b), and has been fitted in previous published works with as few as 13 (Tarko, 2020). Nevertheless, the small number of conflicts is a serious impediment to the level of detail reachable by the safety analysis; with more data, it could have been possible to fit several Lomax distributions, investigating, for example, the safety of pedestrians and bicyclists separately, or focusing on single sub-areas of the study area, potentially highlighting local issues which did not emerge with the more aggregate analysis that was carried out.

In addition, no crash data were available for the study area. Validating the conflict analysis results with historical crash data, by comparing them with the estimated number of crashes in the Before Scenario, would certainly strengthen the argument on the plausibility of the results made in Section 5.2, and this would be especially welcome for case studies with only few data available for distribution fitting. It is, however, worth pointing out that it is common in EVT conflict-based before-after analysis not to validate the results of the models with crash data (Zheng et al., 2018b; Zheng and Sayed, 2019), due to the difficulty to retrieve this kind of data and to the inherent scarcity of crashes, especially for small study areas like the one analyzed in this paper.

Finally, another relevant limitation was the lack of a control site to account for possible confounding factors, which is an effective way to strengthen the finding of before-after analyses (Autey et al., 2012; El-Basyouny and Sayed, 2011; Reyad et al., 2017). Although this practice significantly increases the resources needed to perform the analysis, and it may not be compatible with the time and budget constraints of real-world practices, the present case study is a relevant example that, if possible, a control site should be analyzed, as unexpected and unexpectable conditions may impact the After Scenario, such as, in this case, COVID-19 pandemic. It is reasonable to affirm that the effect of the pandemic was limited, since the main attractors in the study area (i.e., the kindergarten and the bar) were regularly operational, but it is not possible to demonstrate that definitely. In particular, it could be possible that COVID-19 contributed, at least in part, to the overall reduction of traffic volumes in the study area (Section 5.1.1).

## 6.3 Practical takeaways and prescriptions

Keeping in mind that the main objective of the present case study was to illustrate an application of the proposed methodology in a real-world scenario, rather than debating the safety of the specific shared space analyzed, the limitations pointed out in Section 6.2 can be regarded in a positive light, as they provided two main practical takeaways.

First, the choice of survey duration is crucial and should be longer for those case studies in which the vehicles and VRU volumes are small. The conflict analysis of the Before Scenario should be performed immediately after carrying out the survey, so that, if not enough conflicts are detected, it is still possible to perform a supplementary survey before the physical intervention is implemented. The evaluation of whether the number of observed conflicts is "enough" should also account for the possibility that, thanks to the intervention and given equal survey duration, their number may decrease significantly in the After Scenario. It should be noted that if the urban design intervention is effective in improving safety, the number of observed conflicts in the After Scenario may be drastically reduced to the point that it may be unfeasible to produce a crash estimation as reliable as in the Before Scenario, within a reasonably long observation period. However, that would be somewhat evidence of a relevant safety improvement.

Second, one or more additional sites should be surveyed to control the possible presence of confounding factors. They should have characteristics comparable to those of the study area, and their location should be close enough to the study area to be a somewhat representative reference, but far enough not to be influenced by the implemented urban design intervention. In any case, the before-after surveys must be planned to reduce as much as possible the presence of confounding factors (e.g., carrying out the surveys at the same time-of-the-day, in the same weekday, in the same season, and with the same weather).

# 7. Conclusion

This work presented an innovative methodology for before-after safety evaluation of shared space implementations. The procedure, which is structured to be suitable for real-world use by practitioners, aims at filling a gap in the literature, as before-after analyses of shared space are normally limited to spatial behavior investigations, without necessarily providing an unequivocal quantification of safety benefits for vulnerable road users.

In the proposed approach, which is illustrated here with a real-world case study located in Italy, traffic conflicts between vehicles and VRUs are identified by means of a surrogate measure of safety called TTAC. Under the assumption of a probabilistic relationship between conflicts and crashes, the expected number of crashes before and after the implementation of a shared space is retrieved by fitting Extreme Value distributions (Lomax type) with the collected TTAC values. The procedure is rather flexible, as alternative surrogate measures of safety and EVT distributions can be used. In general, it can be applied to evaluate the VRU safety of any urban design intervention.

With respect to the specific case study analyzed, the analysis showed a relevant safety improvement in the After Scenario, with an estimated 53% reduction in crash risk. Nevertheless, spatial behavior analysis highlighted some possible issues (e.g., there is still some segregation between road users, pedestrians tend not to comply with the new crossing), which may have to some extent hampered the safety gains.

The case study analyzed had some shortcomings, most notably a limited number of observed conflicts. While this allowed to highlight some practical issues that may arise in the real world and to provide useful takeaways, future research on this topic would benefit by the application of the proposed methodology to larger case-studies, both in terms of area and observation period. In particular, it could be possible to: (i) provide a more detailed analysis, distinguishing between different VRU classes and sub-areas of the shared space; (ii) address potential confounding factors, by surveying a control site; (iii) compare the results obtained with alternative EVT distributions which require larger fitting datasets.

The present methodology provides a safety evaluation based on the before-after expected number of crashes, without giving information on crash outcomes. Further research will address this issue, testing existing and novel ways to weigh conflict nearness and potential outcome severity (e.g., Extended delta-V method, Laureshyn et al., 2017).

# References

Allen, B.L., Shin, B.T., Cooper, P., 1978. Analysis of Traffic Conflicts and Collisions. Transp. Res. Rec. 67–74.

Alozi, A.R., Hussein, M., 2022. Evaluating the safety of autonomous vehicle–pedestrian interactions: An extreme value theory approach. Anal. Methods Accid. Res. 35, 100230. https://doi.org/10.1016/j.amar.2022.100230

Anciaes, P., Di Guardo, G., Jones, P., 2020. Factors explaining driver yielding behaviour towards pedestrians at courtesy crossings. Transp. Res. Part F Traffic Psychol. Behav. 73, 453–469. https://doi.org/10.1016/j.trf.2020.07.006

Arun, A., Haque, M.M., Bhaskar, A., Washington, S., 2022. Transferability of multivariate extreme value models for safety assessment by applying artificial intelligence-based video analytics. Accid. Anal. Prev. 170, 106644. https://doi.org/10.1016/j.aap.2022.106644

Arun, A., Haque, M.M., Bhaskar, A., Washington, S., Sayed, T., 2021a. A systematic mapping review of surrogate safety assessment using traffic conflict techniques. Accid. Anal. Prev. 153, 106016. https://doi.org/10.1016/J.AAP.2021.106016

Arun, A., Haque, M.M., Washington, S., Sayed, T., Mannering, F., 2021b. A systematic review of traffic conflict-based safety measures with a focus on application context. Anal. Methods Accid. Res. 32, 100185. https://doi.org/10.1016/j.amar.2021.100185

Autey, J., Sayed, T., Zaki, M.H., 2012. Safety evaluation of right-turn smart channels using automated traffic conflict analysis. Accid. Anal. Prev. 45, 120–130. https://doi.org/10.1016/j.aap.2011.11.015

Batista, M., Friedrich, B., 2022a. Investigating spatial behaviour in different types of shared space. Transp. Res. Procedia 60, 44–51. https://doi.org/10.1016/j.trpro.2021.12.007

Batista, M., Friedrich, B., 2022b. Analysing the influence of a farmers' market on spatial behaviour in shared spaces. J. Urban Des. https://doi.org/10.1080/13574809.2022.2042228

Borsos, A., 2021. Application of Bivariate Extreme Value models to describe the joint behavior of temporal and speed related surrogate measures of safety. Accid. Anal. Prev. 159, 106274. https://doi.org/10.1016/j.aap.2021.106274

Borsos, A., Farah, H., Laureshyn, A., Hagenzieker, M., 2020. Are collision and crossing course surrogate safety indicators transferable? A probability based approach using extreme value theory. Accid. Anal. Prev. 143, 105517. https://doi.org/10.1016/J.AAP.2020.105517

Campbell, K.L., Joksch, H.C., Green, P.E., 1996. A bridging analysis for estimating the benefits of active safety technologies. University of Michigan, Transportation Research Institute, Ann Arbor, MI.

Cavadas, J., Azevedo, C.L., Farah, H., Ferreira, A., 2020. Road safety of passing maneuvers: a bivariate extreme value theory approach under non-stationary conditions. Accid. Anal. Prev. 134, 105315. https://doi.org/10.1016/j.aap.2019.105315

Chaudhari, A., Gore, N., Arkatkar, S., Joshi, G., Pulugurtha, S., 2021. Exploring pedestrian surrogate safety measures by road geometry at midblock crosswalks: A perspective under mixed traffic conditions. IATSS Res. 45, 87–101. https://doi.org/10.1016/j.iatssr.2020.06.001

Chen, P., Zeng, W., Yu, G., Wang, Y., 2017. Surrogate Safety Analysis of Pedestrian-Vehicle Conflict at Intersections Using Unmanned Aerial Vehicle Videos. J. Adv. Transp. 2017. https://doi.org/10.1155/2017/5202150

Clarke, E., 2006. Shared Space - The alternative a approach to calming traffic. Traffic Eng. Control 47, 290.

Danaf, M., Sabri, A., Abou-Zeid, M., Kaysi, I., 2020. Pedestrian–vehicular interactions in a mixed street environment. Transp. Lett. 12, 87–99. https://doi.org/10.1080/19427867.2018.1525821

Davis, G.A., Hourdos, J., Xiong, H., Chatterjee, I., 2011. Outline for a causal model of traffic conflicts and crashes. Accid. Anal. Prev. 43, 1907–1919. https://doi.org/10.1016/j.aap.2011.05.001

El-Basyouny, K., Sayed, T., 2013. Safety performance functions using traffic conflicts. Saf. Sci. 51, 160–164. https://doi.org/10.1016/J.SSCI.2012.04.015

El-Basyouny, K., Sayed, T., 2011. A full Bayes multivariate intervention model with random parameters among matched pairs for before-after safety evaluation. Accid. Anal. Prev. 43, 87–94. https://doi.org/10.1016/j.aap.2010.07.015

Ezzati Amini, R., Yang, K., Antoniou, C., 2022. Development of a conflict risk evaluation model to assess pedestrian safety in interaction with vehicles. Accid. Anal. Prev. 175, 106773. https://doi.org/10.1016/J.AAP.2022.106773

Fagerland, M.W., Lydersen, S., Laake, P., 2013. The McNemar test for binary matched-pairs data: Mid-p and asymptotic are better than exact conditional. BMC Med. Res. Methodol. 13, 91. https://doi.org/10.1186/1471-2288-13-91

Farah, H., Azevedo, C.L., 2017. Safety analysis of passing maneuvers using extreme value theory. IATSS Res. 41, 12–21. https://doi.org/10.1016/j.iatssr.2016.07.001

Fu, C., Sayed, T., Zheng, L., 2021. Multi-type Bayesian hierarchical modeling of traffic conflict extremes for crash estimation. Accid. Anal. Prev. 160, 106309. https://doi.org/10.1016/j.aap.2021.106309

Fu, C., Sayed, T., Zheng, L., 2020. Multivariate Bayesian hierarchical modeling of the non-stationary traffic conflict extremes for crash estimation. Anal. Methods Accid. Res. 28, 100135. https://doi.org/10.1016/j.amar.2020.100135

Gastaldi, M., Orsini, F., Gecchele, G., Rossi, R., 2021. Safety analysis of unsignalized intersections: a bivariate extreme value approach. Transp. Lett. 13, 209–218. https://doi.org/10.1080/19427867.2020.1861503

Glauz, W.D., Migletz, D.J., 1980. APPLICATION OF TRAFFIC CONFLICT ANALYSIS AT INTERSECTIONS. NCHRP Rep.

Gruden, C., Otković, I.I., Šraml, M., 2022. Pedestrian safety at roundabouts: a comparison of the behavior in Italy and Slovenia. Transp. Res. Procedia 60, 528–535. https://doi.org/10.1016/j.trpro.2021.12.068

Guo, F., Klauer, S.G., Hankey, J.M., Dingus, T.A., 2010. Near Crashes as Crash Surrogate for Naturalistic Driving Studies: https://doi.org/10.3141/2147-09 66–74. https://doi.org/10.3141/2147-09

Hamilton-Baillie, B., 2008. Shared space: Reconciling people, places and traffic. Built Environ. 34, 161–181. https://doi.org/10.2148/benv.34.2.161

Haque, M.M., Oviedo-Trespalacios, O., Sharma, A., Zheng, Z., 2021. Examining the driver-pedestrian interaction at pedestrian crossings in the connected environment: A Hazard-based duration modelling approach. Transp. Res. Part A Policy Pract. 150, 33–48. https://doi.org/10.1016/J.TRA.2021.05.014

Hauer, E., 1982. Traffic conflicts and exposure. Accid. Anal. Prev. 14, 359–364. https://doi.org/10.1016/0001-4575(82)90014-8

Hayward J.C., 1972. Near miss determination through use of a scale of danger. Highw. Res. Rec. https://doi.org/TTSC 7115

Hirsch, L., Mackie, H., Crombie, C., Bolton, L., Wilson, N., Cornille, Z., 2022. Road user interaction changes following street improvements from Te Ara Mua – Future Streets: A case study. J. Transp. Heal. 25, 101384. https://doi.org/10.1016/j.jth.2022.101384

Hydén, C., 1987. The development of a method for traffic safety evaluation:the Swedish traffic conflicts technique. Bull. Lund Inst. od Technol. 70.

Jiang, X., Wang, W., Bengler, K., 2015. Intercultural Analyses of Time-to-Collision in Vehicle-Pedestrian Conflict on an Urban Midblock Crosswalk. IEEE Trans. Intell. Transp. Syst. 16, 1048–1053. https://doi.org/10.1109/TITS.2014.2345555

Johnsson, C., Laureshyn, A., Dágostino, C., 2021. Validation of surrogate measures of safety with a focus on bicyclist–motor vehicle interactions. Accid. Anal. Prev. 153, 106037. https://doi.org/10.1016/j.aap.2021.106037

Johnsson, C., Laureshyn, A., De Ceunynck, T., 2018. In search of surrogate safety indicators for vulnerable road users: a review of surrogate safety indicators. Transp. Rev. 38, 1–21. https://doi.org/10.1080/01441647.2018.1442888

Kaparias, I., Bell, M., Dong, W., Sastrawinata, A., Singh, A., Wang, X., Mount, B., 2013. Analysis of pedestrian-vehicle traffic conflicts in street designs with elements of shared space. Transp. Res. Rec. 21–30. https://doi.org/10.3141/2393-03

Kaparias, I., Bell, M.G.H., Biagioli, T., Bellezza, L., Mount, B., 2015. Behavioural analysis of interactions between pedestrians and vehicles in street designs with elements of shared space. Transp. Res. Part F Traffic Psychol. Behav. 30, 115–127. https://doi.org/10.1016/j.trf.2015.02.009

Kaparias, I., Bell, M.G.H., Greensted, J., Cheng, S., Miri, A., Taylor, C., Mount, B., 2010. Development and implementation of a vehicle-pedestrian conflict analysis method: Adaptation of a vehicle-vehicle technique. Transp. Res. Rec. 75–82. https://doi.org/10.3141/2198-09

Kaparias, I., Hirani, J., Bell, M.G.H., Mount, B., 2016. Pedestrian gap acceptance behavior in street designs with elements of shared space. Transp. Res. Rec. 2586, 17–27. https://doi.org/10.3141/2586-03

Karndacharuk, A., Wilson, D., Dunn, R., 2013. Analysis of pedestrian performance in shared-space environments. Transp. Res. Rec. 1–11. https://doi.org/10.3141/2393-01

Karndacharuk, A., Wilson, D.J., Dunn, R., 2014a. A Review of the Evolution of Shared (Street) Space Concepts in Urban Environments. Transp. Rev. 34, 190–220. https://doi.org/10.1080/01441647.2014.893038

Karndacharuk, A., Wilson, D.J., Dunn, R.C.M., 2014b. Safety performance study of shared pedestrian and vehicle space in New Zealand. Transp. Res. Rec. 2464, 1–10. https://doi.org/10.3141/2464-01

Kizawi, A., Borsos, A., 2021. A Literature review on the conflict analysis of vehicle-pedestrian interactions. Acta Tech. Jaurinensis 14, 599–611. https://doi.org/10.14513/ACTATECHJAUR.00601

Kovaceva, J., Nero, G., Bärgman, J., Dozza, M., 2019. Drivers overtaking cyclists in the real-world: Evidence from a naturalistic driving study. Saf. Sci. 119, 199–206. https://doi.org/10.1016/j.ssci.2018.08.022

Laureshyn, A., De Ceunynck, T., Karlsson, C., Svensson, Å., Daniels, S., 2017. In search of the severity dimension of traffic events: Extended Delta-V as a traffic conflict indicator. Accid. Anal. Prev. 98, 46–56. https://doi.org/10.1016/j.aap.2016.09.026

Laureshyn, A., Svensson, Å., Hydén, C., 2010. Evaluation of traffic safety, based on micro-level behavioural data: Theoretical framework and first implementation. Accid. Anal. Prev. 42, 1637–1646. https://doi.org/10.1016/j.aap.2010.03.021

Lee, H., Kim, S.N., 2019. Shared space and pedestrian safety: Empirical evidence from Pedestrian Priority Street projects in Seoul, Korea. Sustain. 11. https://doi.org/10.3390/su11174645

Lenard, J., Welsh, R., Danton, R., 2018. Time-to-collision analysis of pedestrian and pedal-cycle accidents for the development of autonomous emergency braking systems. Accid. Anal. Prev. 115, 128–136. https://doi.org/10.1016/J.AAP.2018.02.028

McNemar, Q., 1947. Note on the sampling error of the difference between correlated proportions or percentages. Psychometrika 12, 153–157. https://doi.org/10.1007/BF02295996

Miani, M., Dunnhofer, M., Micheloni, C., Marini, A., Baldo, N., 2022. Young drivers' pedestrian anti-collision braking operation data modelling for ADAS development. Transp. Res. Procedia 60, 432–439. https://doi.org/10.1016/j.trpro.2021.12.056

Miani, M., Dunnhofer, M., Micheloni, C., Marini, A., Baldo, N., 2021. Surrogate safety measures prediction at multiple timescales in v2p conflicts based on gated recurrent unit. Sustain. 13, 9681. https://doi.org/10.3390/su13179681

Nadimi, N., Ragland, D.R., Mohammadian Amiri, A., 2020. An evaluation of time-to-collision as a surrogate safety measure and a proposal of a new method for its application in safety analysis. Transp. Lett. 12, 491–500. https://doi.org/10.1080/19427867.2019.1650430

Nasernejad, P., Sayed, T., Alsaleh, R., 2021. Modeling pedestrian behavior in pedestrian-vehicle near misses: A continuous Gaussian Process Inverse Reinforcement Learning (GP-IRL) approach. Accid. Anal. Prev. 161, 106355. https://doi.org/10.1016/J.AAP.2021.106355

Navarro, B., Miranda-Moreno, L., Saunier, N., Labbe, A., Fu, T., 2022. Do stop-signs improve the safety for all road users? A before-after study of stop-controlled intersections using video-based trajectories and surrogate measures of safety. Accid. Anal. Prev. 167, 106563. https://doi.org/10.1016/j.aap.2021.106563

Orsini, F., Gecchele, G., Gastaldi, M., Rossi, R., 2020. Large-scale road safety evaluation using extreme value theory. IET Intell. Transp. Syst. 14, 1004–1012. https://doi.org/10.1049/iet-its.2019.0633

Orsini, F., Gecchele, G., Gastaldi, M., Rossi, R., 2019. Collision prediction in roundabouts: a comparative study of extreme value theory approaches. Transp. A Transp. Sci. 15, 556–572. https://doi.org/10.1080/23249935.2018.1515271

Orsini, F., Zarantonello, L., Costa, A., Rossi, R., Montagnese, S., 2022. Driving simulator performance worsens after the Spring transition to Daylight Saving Time. iScience 25, 104666. https://doi.org/10.1016/j.isci.2022.104666

Pascucci, F., Rinke, N., Schiermeyer, C., Friedrich, B., Berkhahn, V., 2015. Modeling of shared space with multi-modal traffic using a multi-layer social force approach. Transp. Res. Procedia 10, 316–326. https://doi.org/10.1016/j.trpro.2015.09.081

Pembury Smith, M.Q.R., Ruxton, G.D., 2020. Effective use of the McNemar test. Behav. Ecol. Sociobiol. 74. https://doi.org/10.1007/s00265-020-02916-y

Reyad, P., Sacchi, E., Ibrahim, S., Sayed, T., 2017. Traffic conflict-based before-After study with use of comparison groups and the empirical Bayes method. Transp. Res. Rec. 2659, 15–24. https://doi.org/10.3141/2659-02

Rinke, N., Schiermeyer, C., Pascucci, F., Berkhahn, V., Friedrich, B., 2017. A multi-layer social force approach to model interactions in shared spaces using collision prediction. Transp. Res. Procedia 25, 1249–1267. https://doi.org/10.1016/j.trpro.2017.05.144

Rossi, R., Orsini, F., Tagliabue, M., Di Stasi, L.L., De Cet, G., Gastaldi, M., 2021. Evaluating the impact of real-time coaching programs on drivers overtaking cyclists. Transp. Res. Part F Traffic Psychol. Behav. 78, 74–90. https://doi.org/10.1016/j.trf.2021.01.014

Ruiz-Apilánez, B., Karimi, K., García-Camacha, I., Martín, R., 2017. Shared space streets: Design, user perception and performance. Urban Des. Int. 22, 267–284. https://doi.org/10.1057/s41289-016-0036-2

Salamati, K., Schroeder, B., Rouphail, N.M., Cunningham, C., Long, R., Barlow, J., 2011. Development and implementation of conflict-based assessment of pedestrian safety to evaluate accessibility of complex intersections. Transp. Res. Rec. 148–155. https://doi.org/10.3141/2264-17

Sayed, T., Zein, S., 1999. Traffic conflict standards for intersections. Transp. Plan. Technol. 22:4, 309–323. https://doi.org/10.1080/03081069908717634

Songchitruksa, P., Tarko, A.P., 2006. The extreme value theory approach to safety estimation. Accid. Anal. Prev. 38, 811–822. https://doi.org/10.1016/j.aap.2006.02.003

Tahir, H. Bin, Haque, M.M., Yasmin, S., King, M., 2022. A simulation-based empirical bayes approach: Incorporating unobserved heterogeneity in the before-after evaluation of engineering treatments. Accid. Anal. Prev. 165, 106527. https://doi.org/10.1016/j.aap.2021.106527

Tarko, A.P., 2021. A unifying view on traffic conflicts and their connection with crashes. Accid. Anal. Prev. 158, 106187. https://doi.org/10.1016/j.aap.2021.106187

Tarko, A.P., 2020. Analyzing road near departures as failure-caused events. Accid. Anal. Prev. 142, 105536. https://doi.org/10.1016/j.aap.2020.105536

Tarko, A.P., 2019. Measuring road safety with surrogate events, Measuring Road Safety with Surrogate Events. Elsevier. https://doi.org/10.1016/C2016-0-00255-3

Tarko, A.P., 2018a. Chapter 17. Surrogate Measures of Safety, in: Safe Mobility: Challenges, Methodology and Solutions. pp. 383–405. https://doi.org/10.1108/S2044-994120180000011019

Tarko, A.P., 2018b. Estimating the expected number of crashes with traffic conflicts and the Lomax Distribution – A theoretical and numerical exploration. Accid. Anal. Prev. 113, 63–73. https://doi.org/10.1016/j.aap.2018.01.008

Tarko, A.P., Lizarazo, C.G., 2021. Validity of failure-caused traffic conflicts as surrogates of rear-end collisions in naturalistic driving studies. Accid. Anal. Prev. 149, 105863. https://doi.org/10.1016/j.aap.2020.105863

Trifunović, A., Timmermann, C., Friedrich, B., Berkhahn, V., 2021. Implications of Converting Low Capacity Intersection Adjacent to Park Into a Shared Space, in: Transportation Research Board 100th Annual Meeting. Washington D.C., p. TRBAM-21-02466.

Vasudevan, V., Agarwala, R., Tiwari, A., 2022. LiDAR-Based Vehicle–Pedestrian Interaction Study on Midblock Crossing Using Trajectory-Based Modified Post-Encroachment Time: https://doi.org/10.1177/03611981221083295 036119812210832. https://doi.org/10.1177/03611981221083295

Wang, C., Xu, C., Xia, J., Qian, Z., Lu, L., 2018. A combined use of microscopic traffic simulation and extreme value methods for traffic safety evaluation. Transp. Res. Part C Emerg. Technol. 90, 281–291. https://doi.org/10.1016/j.trc.2018.03.011

Zangenehpour, S., Strauss, J., Miranda-Moreno, L.F., Saunier, N., 2016. Are signalized intersections with cycle tracks safer? A case-control study based on automated surrogate safety analysis using video data. Accid. Anal. Prev. 86, 161–172. https://doi.org/10.1016/j.aap.2015.10.025

Zhang, S., Abdel-Aty, M., Wu, Y., Zheng, O., 2020. Modeling pedestrians' near-accident events at signalized intersections using gated recurrent unit (GRU). Accid. Anal. Prev. 148, 105844. https://doi.org/10.1016/j.aap.2020.105844

Zhang, Y., Yao, D., Qiu, T.Z., Peng, L., 2012. Pedestrian safety analysis in mixed traffic conditions using video data. IEEE Trans. Intell. Transp. Syst. 13, 1832–1844. https://doi.org/10.1109/TITS.2012.2210881

Zhang, Y., Yao, D., Qiu, T.Z., Peng, L., 2011. Vehicle-pedestrian interaction analysis in mixed traffic condition. ICTIS 2011 Multimodal Approach to Sustain. Transp. Syst. Dev. - Information, Technol. Implement. - Proc. 1st Int. Conf. Transp. Inf. Saf. 552–559. https://doi.org/10.1061/41177(415)70

Zhao, Y., Miyahara, T., Mizuno, K., Ito, D., Han, Y., 2021. Analysis of car driver responses to avoid car-to-cyclist perpendicular collisions based on drive recorder data and driving simulator experiments. Accid. Anal. Prev. 150, 105862. https://doi.org/10.1016/j.aap.2020.105862

Zheng, L., Ismail, K., Meng, X., 2014. Freeway safety estimation using extreme value theory approaches: A comparative study. Accid. Anal. Prev. 62, 32–41. https://doi.org/10.1016/j.aap.2013.09.006

Zheng, L., Ismail, K., Sayed, T., Fatema, T., 2018a. Bivariate extreme value modeling for road safety estimation. Accid. Anal. Prev. 120, 83–91. https://doi.org/https://doi.org/10.1016/j.aap.2018.08.004

Zheng, L., Sayed, T., 2019. Application of Extreme Value Theory for Before-After Road Safety Analysis. Transp. Res. Rec. 0361198119841555. https://doi.org/10.1177/0361198119841555

Zheng, L., Sayed, T., Essa, M., 2019. Bayesian hierarchical modeling of the non-stationary traffic conflict extremes for crash estimation. Anal. Methods Accid. Res. 23. https://doi.org/10.1016/j.amar.2019.100100

Zheng, L., Sayed, T., Tageldin, A., 2018b. Before-after safety analysis using extreme value theory: A case of left-turn bay extension. Accid. Anal. Prev. 121, 258–267. https://doi.org/https://doi.org/10.1016/j.aap.2018.09.023

Zhu, H., Almukdad, A., Iryo-Asano, M., Alhajyaseen, W.K.M., Nakamura, H., Zhang, X., 2021. A novel agent-based framework for evaluating pedestrian safety at unsignalized mid-block crosswalks. Accid. Anal. Prev. 159, 106288. https://doi.org/10.1016/J.AAP.2021.106288